\documentclass[sigconf]{acmart}

\usepackage{algorithmic, algorithm, framed, xcolor, colortbl, subfigure, textcomp, pdfpages, xspace}
\usepackage{booktabs}
\usepackage{multirow}
\usepackage{graphicx} % omit 'demo' for real document

\newtheorem{Theorem}{Theorem}

\usepackage[british]{babel} 
\usepackage[colorinlistoftodos, textwidth=20mm, textsize=footnotesize]{todonotes}

\AtBeginDocument{%
  \providecommand\BibTeX{{%
    \normalfont B\kern-0.5em{\scshape i\kern-0.25em b}\kern-0.8em\TeX}}}

\setcopyright{acmcopyright}
\acmConference{Manuscript Draft}{April 2020}{Ume\aa\ University}
\renewcommand\footnotetextcopyrightpermission[1]{} 
\begin{document}

\newcommand{\systemname}{E-HPC\xspace}
\newcommand{\systemone}{HPC2n\xspace} %HPC2n
\newcommand{\systemtwo}{UPPMAX\xspace} %UPPMAX
%%
%% The "title" command has an optional parameter,
%% allowing the author to define a "short title" to be used in page headers.
\title{ASA - The Adaptive Scheduling Algorithm}

%%
%% The "author" command and its associated commands are used to define
%% the authors and their affiliations.
%% Of note is the shared affiliation of the first two authors, and the
%% "authornote" and "authornotemark" commands
%% used to denote shared contribution to the research.
\author{Abel Souza}
%\authornote{Both authors contributed equally to this research.}
%\orcid{1234-5678-9012}
%\author{G.K.M. Tobin}
%\authornotemark[1]
%\email{webmaster@marysville-ohio.com}
\affiliation{%
  \institution{University of Massachusetts Amherst}
  %\streetaddress{P.O. Box 1212}
  %\city{Dublin}
  %\state{Sweden}
  %\postcode{43017-6221}
}

\author{Kristiaan Pelckmans}
\affiliation{%
  \institution{Uppsala University}
  %\streetaddress{1 Th{\o}rv{\"a}ld Circle}
  %\city{Hekla}
  %\country{Sweden}
  }

\author{Devarshi Ghoshal, Lavanya Ramakrishnan}
\affiliation{%
  \institution{Lawrence Berkeley National Lab.}
  %\city{Berkeley}
  %\state{California}
  %\country{USA}
}

\author{Johan Tordsson}
\affiliation{%
 \institution{Elastisys and Ume\aa\ University}
 %\streetaddress{Rono-Hills}
 %\city{Doimukh}
 %\state{Arunachal Pradesh}
 %\country{Sweden}
 }

%%
%% By default, the full list of authors will be used in the page
%% headers. Often, this list is too long, and will overlap
%% other information printed in the page headers. This command allows
%% the author to define a more concise list
%% of authors' names for this purpose.
\renewcommand{\shortauthors}{A Souza, et al.}

%%
%% The abstract is a short summary of the work to be presented in the
%% article.
\begin{abstract}
In High Performance Computing (HPC) infrastructures, the control of resources by batch systems can lead to prolonged queue waiting times and adverse effects on the overall execution times of applications, particularly in data-intensive and low-latency workflows where efficient processing hinges on resource planning and timely allocation. Allocating the maximum capacity upfront ensures the fastest execution but results in spare and idle resources, extended queue waits, and costly usage. Conversely, dynamic allocation based on workflow stage requirements optimizes resource usage but may negatively impact the total workflow makespan. To address these issues, we introduce ASA, the Adaptive Scheduling Algorithm. ASA is a novel, convergence-proven scheduling technique that minimizes jobs inter-stage waiting times by estimating the queue waiting times to proactively submit resource change requests ahead of time. It strikes a balance between exploration and exploitation, considering both learning (waiting times) and applying learnt insights. Real-world experiments over two supercomputers centers with scientific workflows demonstrate ASA's effectiveness, achieving near-optimal resource utilization and accuracy, with up to 10\% and 2\% reductions in average workflow queue waiting times and makespan, respectively.
\end{abstract}

%%
%% Keywords. The author(s) should pick words that accurately describe
%% the work being presented. Separate the keywords with commas.
%\keywords{datasets, neural networks, gaze detection, text tagging}

%%
%% This command processes the author and affiliation and title
%% information and builds the first part of the formatted document.
\maketitle

\section{INTRODUCTION}

\begin{figure}[t] \centering
	\includegraphics[width=\linewidth]{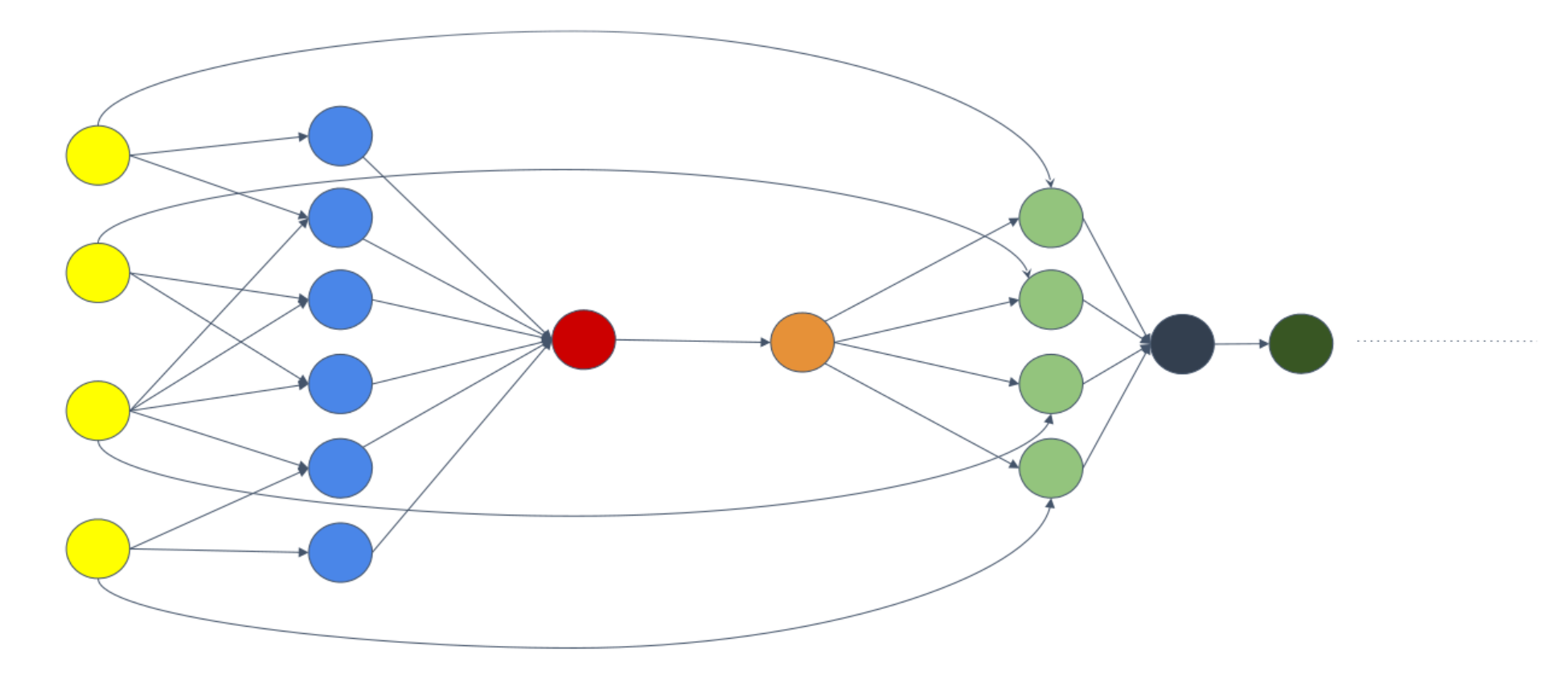}
	\DeclareGraphicsExtensions.
	\caption{Excerpt from the Montage scientific workflow, an image mosaic software employed by NASA~\cite{berriman2004montage}. Different colors in the graph represent distinct sets of tasks within a stage. Outputs generated in each stage serve as inputs for subsequent stages, ultimately culminating in the final result.}
	\label{fig:montage}
\end{figure}

Large scale experiments model different aspects of nature such as weather forecasting, drug discovery, fluid dynamics, and many other scientific endeavours.
Higher resolution sensors have been generating an ever larger amount of data, usually processed over large and complex computing infrastructures such as High Performance (HPC) and cloud computing datacenters. 
Due to its complexity with modeling and handling great amounts of data, such time consuming scientific campaigns are organized in independent data pipelines, known as \textit{scientific workflows} \cite{taylor2007workflows}. 
Figure \ref{fig:montage} exemplifies parts of the Montage Workflow, an image mosaic engine \cite{montage}.
A scientific workflow is composed of sequentially interconnected stages (the different colors in Figure \ref{fig:montage}), where each stage is responsible for a specific set of tasks inside the overall application data flow.
Scientific workflows are not only common in HPC centres, but also virtually in every sector of industry and academia, where they are used for analyzing and correlating data for predictions and decision support.

%Batch calculations are nowadays not only common in HPC, but are also used in virtually every sector to analyze and correlate data for predictions and decision support.

Intrinsically, a stage in a workflow structure describes its scalability and the amount of resources required to perform all of its tasks.
%Such data pipeline stages are managed by a workflow management system (WMS).
To ensure acceptable task performance during workflow execution is the responsibility of the developers and the workflow management systems (WMS).
When time to scale the developed workflow comes, users make use of HPC infrastructures.
However, HPC platforms are primarily designed to support monolithic applications and provide a static allocation scheduling model i.e., the resource allocation is fixed throughout the entire job lifespan \cite{jha2014tale, reuther2018scalable, 2013omega, Compres:2016:IAE:2966884.2966917}.
This methodology guarantees good performance, however it results in fragmentation and lower datacenter efficiency due to under-utilization.
It also hinders the development of newer scheduling strategies needed in dynamic computational models, like data intensive and streaming workflows, increasingly used for conducting online and in-situ experiments \cite{deelman2018future}.
These problems are likely to exacerbate with highly dynamic workflows in the next-generation exascale systems \cite{bergman2008exascale}, expected to have applications issuing and orchestrating thousands of simultaneous processes \cite{castain2018pmix}.
With increasing use of workflows to process big amounts of data, a closer integration between the WMS and the datacenter resource manager (RM) is of vital importance for meeting scientific application constraints, like placement, resource isolation and control, turnaround times, and overall datacenter efficiency \cite{deelman2018future,Compres:2016:IAE:2966884.2966917,asch2018big}.
%on the amount of resources used dynamically at runtime 

%new models and architectures enabling workflow orchestration and adaptation to infrastructural dynamics, like amount of resource changes, workflow processes misbehaves, are key not only to current workflows, but also to future data intensive workflows.
%In order to simultaneously improve workflow makespan, resource planning and utilization, a tighter integration with the resource manager is needed.
%\comm{Contribution (i), unfortunately, is not novel. This idea has been proposed before (e.g., in E-HPC: A Library for Elastic Resource Management in HPC Environments" by Fox et al). In fact, the work by Zhang et al. (reference [26]) goes much further and fuses workflow levels together for better performance. Essentially all aspects of this scheduling problem as described in this work has been previously described/published before (e.g., the notion of asking more time in case a job starts too early is the "leeway" in [26]).}

In this paper we propose ASA: the Adaptive Scheduling Algorithm.
Leveraging conceptual ideas from distributed operating systems \cite{2011mesos}, ASA decouples application development and scheduling planning from resource management.
Packaged as a library, ASA presents applications with resources from multiple job allocations as one global pool of resources.
This allows workflow management systems to be fault-tolerant, elastic, besides enabling the use of new scheduling strategies. 
%enabling fault-tolerant and elastic distributed systems to easily be built and run effectively. 
ASA pro-actively estimates the waiting time for coming stages in a workflow during the currently executing stage to improve workflow turnaround times. 
In this way, ASA not only optimizes total resource usage, but it also reduces the total workflow makespan.
For estimating how an user waits in the queue, we designed a simple Reinforcement Learning (RL) algorithm, which can adapt to the current state of a queue.
%Besides that, estimate the likelihood of the queue waiting times for each type of resource request.
%We study a novel statistical strategy for pro-actively placing job requests to minimize waiting times, idle resources, and to reduce workflow's total runtime, all managed by ASA.
This amounts to both learning (the waiting times), and acting on what is learnt thus far, and amounts hence to a realization of the exploration-exploitation trade-off.
Experiments with real workflows in real supercomputers show that ASA achieves a middle ground between the two aforementioned ways for resource allocation: lower total turnaround times, with near optimal resource utilization.

%The rest of this paper is organized as follows.
%In Section~\ref{sec:background}, we describe in more details the characteristics of scientific workflows, their scheduling trade-offs, past work, and challenges.
%In Section~\ref{asa:architecture}, we present ASA, an architecture and algorithm for resource orchestration, management, and planning.
%Experiments, evaluations, various analyses, and discussion follow in Sections~\ref{sec:results} and \ref{sec:disc}.
%Finally, we present our conclusions in Section~\ref{sec:conc}.
\section{Background and Related Work}\label{sec:background}

The nature of scientific models is very complex and thus projects are often organized in distributed collaborations.
Rather than developing large monolithic applications, scientists use \textit{scientific workflows}: runtime systems for describing and executing applications as pipelined distributed components.
%In order to define, model, and describe the overall software used in such collaborations, 
%Each pipeline describes a specific set of models, instructions, and computational tasks organized in batch calculations, also known as workflow stages.
For example, data modeling, staging, handling, processing, and pre- and post-processing are concrete tasks that may occur before, within, or after these pipeline stages.
%At a high level perspective, these interconnected pipelines composed of independent stages is what defines a \textit{scientific workflow}.
Figure \ref{fig:montage} shows the Montage workflow, an image mosaic application used at NASA \cite{montage} with seven sequential stages, each colored differently.
Edges describe sequential data dependencies, where data outputs of a previous stage are sent to following stages produced at the end of such (previous) stages.
The number of nodes (shown as color circles in Figure \ref{fig:montage}) in a stage describes its scalability: one node means the stage is inherently sequential, using only one available resource (e.g. CPU/core, GPU, etc.), whereas two or more nodes mean parallel stages that may use more than one resource.
Streaming workflows are used in in-situ and online experiments, where all stages run concurrently and data (known as \textit{tuples}) are continuously streamed over the workflow pipeline and processed by each stage as they arrive from predecessor stages \cite{tolosana2011autonomic}.

%Yes, "Workflow" tends to be one of the most abused word in the industry.
%
%People seem to use workflow for all of these things:
%
%#1 Business process modeling and automating platforms - Camunda, BPMN, jBPM etc.
%#2 Assembling together software functions (also called tasks, sometimes) for getting an output data - Airflow, Cadence, CWL is related too.
%#3 Moving one entity (bug, document, deal) through different stages of a state machine, until the entity reaches an end state - Sharepoint workflow, Zoho Orchestly, Comala workflows
%#4 Orchestrating several apps together for automating something, usually as reactions for events - Zapier

%When developers talk about workflows, it's mostly about #2. When business people talk about workflows it's mostly about #1. When managers/leads of a team talk about workflows it's mostly about #3.
%We should accept that Workflow as a word is too ambiguous, and come up with what kind of workflow category the tool belongs to, before branding it as an "one stop solution for all things workflow".

\subsection{Related Work}

Scientific workflows are orchestrated, scheduled, and managed by a WMS, realized by programming language extensions through Application Program Interfaces (API), or by new dialects of common programming languages \cite{amstutz2016common}.
WMS's are used as execution engines for helping users and developers to run, scale, and integrate the distributed components of a workflow. 
Some WMS examples are Apache Taverna \cite{oinn2004taverna}, Kepler \cite{altintas2004kepler}, Pegasus \cite{deelman2004pegasus}, and Tigres \cite{hendrix2016tigres}.
Because these systems do not assume specific runtime behaviors like resources' performance variability, most of them do not support Quality-of-Services (QoS) application requirements.
To overcome such limitations, VGrADS \cite{ramakrishnan2009vgrads} combines resources from different providers into a single virtualized abstraction layer to enable applications with smarter scheduling and fault-tolerant strategies.
Additionally, new tools enable WMS's with stage elasticity, achieving optimal resource expenditures, though with larger makespans \cite{ehpc}.

In environments like HPC centers, jobs have to wait in (priority) queues for resources before starting execution \cite{guim2007job}.
Thus, a natural way to improve total workflow makespan and to enable deadline planning features, is to estimate the queue waiting times. 
For this, three main approaches have been used: (i) simulating scheduling according to the job queue (at certain point in time), (ii) statistical modeling, and (iii) a mix of these two \cite{Sivagnanam:2010:TRS:1838574.1838594}.
Queue simulation (i) is a way to predict waiting time, but can be challenging if estimations needed at runtime do not take into account future (non-deterministic) job submissions from other users, which may degrade predictions.
Although (i) can be used as a baseline for comparisons with more elaborate methods, and though a normal user may have access to the queue statistics for doing so, static methods can be seen as non adaptive if they do not adjust to such queue workload changes.

Traditional Machine Learning models (ii) tend to overfit the wait time because the dataset used for training can rapidly change. 
Its application without understanding the system and its workload does not work well as boundaries and medians tend to produce great over-estimations on the waiting time.
QBets \cite{nurmi2007qbets} used to be a reference system, but it is not in production anymore due to today's workload high dynamicity with sudden changes, well captured with time-series analysis.
QBest predictions were not bounded, and did not take into account variables that affect the job wait time.
QBets' solution ('quantile prediction') is quite different from ASA (see Section \ref{asa:architecture}), as QBet is based on traditional learning, not online learning.
A recent solution is implemented by Karnak \cite{karnark}.
In this work, a large number of variables are used to model the wait time, including seasonal patterns, current system load, queue composition, job geometry, particular user, particular queue, particular group, etc.
A decision tree is used to classify jobs according to different criteria, and then the  resulting bag of jobs are modelled individually, giving better precision.
In a second version, wait times are improved with scheduling simulations.
Although not perfect, this model was shown to achieve much better results than QBets.
Suggestions on the future work of Karnak \cite{ImproveKarnark} point towards the application of neural networks.

\subsection{Scheduling Tradeoffs for Scientific Workflows}

\begin{figure}
    %\centering
    \subfigure[Big Job Strategy]
    {
        \includegraphics[width=\linewidth]{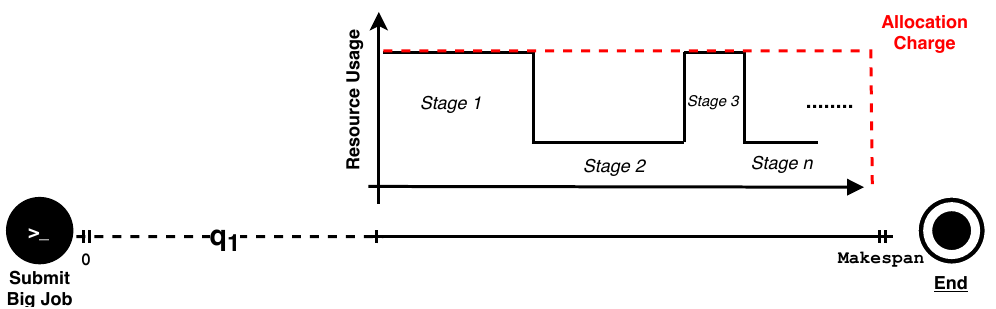}
        \label{fig:bigjob}
    }\\
    \subfigure[Per-stage Strategy]
    {
        \includegraphics[width=\linewidth]{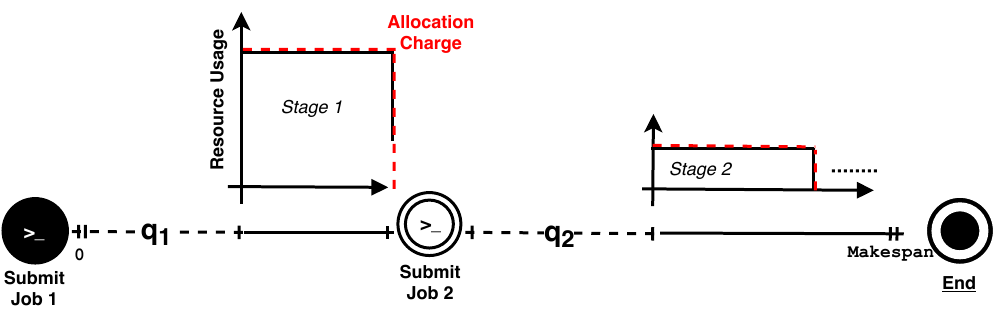}
        \label{fig:per_stage}
    }
    \caption{(a) Big Job vs (b) Per-Stage managed resource allocation strategies in HPC. Fig. \ref{fig:bigjob}: an unique allocation for the entire workflow duration, with single queue waiting time.
 Fig. \ref{fig:per_stage}: per-stage allocations with only as many resources as required by a
 particular stage, with extra inter-stage queue waiting times. Note the differences in makespan and resources charging in each case (summation of area(s) under the dashed red lines).}
    \label{fig:ehpc_overview}
\end{figure}

In this subsection, we formulate scheduling tradeoffs for different strategies when submitting workflow jobs to HPC environments.
We analyze the differences in total resource expenditure and workflow turnaround time.

%Libraries like \systemname \cite{ehpc} are targeted to improve resource usage of workflows, expressed as core-usage. 

%\devarshi{I see you refer to BigJob scheduling strategy in a lot of places, without really defining it first anywhere. Also Fig. 2 somehow should mention BigJob somewhere (may be something like Traditional, BigJob strategy).}

Commonly, users submit scientific workflow jobs to HPC clusters using two different strategies, as shown in Figure \ref{fig:ehpc_overview}.
In Figure~\ref{fig:ehpc_overview}a, a workflow job is submitted as a big allocation (Big Job Strategy in the figure). 
Mathematically, the core-usage ($C$) is defined as $C = n * t$, where $n$ is the number of cores assigned during job execution, and $t$ is the allocated time (often measured in hours, and specified as \textit{core-hours}).
Hence, for a workflow with $s$ stages and each needing time $t_i$ to execute, the total core-hours usage is calculated as the sum of core-hours used by each stage as
\begin{align} \label{eq:ch_nehpc}
C_{Big Job} = n * \sum_{i=1}^{s}{t_i}.
\end{align}
Because stages with different resource requirements are not taken into consideration, the maximum amount of resources $n$ is allocated for the entire duration of the workflow lifespan.
This wastes resources (white areas above the resource usage black, and under the dashed red lines in Figure \ref{fig:ehpc_overview}a), but guarantees lower total workflows' execution times.
Alternatively, users can manually manage the different stages in the workflow by submitting them as multiple sub-jobs.
\systemname is a library that does exactly this, providing elasticity for workflows running over HPC resources \cite{ehpc}. 
Figure \ref{fig:ehpc_overview}b shows this per-stage resource assignment.
The change in the amount of resources occurs at the end of each stage, where a coming stage is assigned with the exact number of cores required for its execution.
Thus $C_{Per-Stage}$, the core-hours usage in a per-stage managed workflow with $s$ stages, is calculated as 
\begin{align}
\label{eq:ch_ehpc}
C_{Per-Stage} = \sum_{i=1}^{s}{(t_i * n_i)},
\end{align}
\noindent where $t_i$ and $n_i$ respectively represent the time and the number of resources needed to execute the $i$-th stage.
Comparing definitions (\ref{eq:ch_nehpc}) and (\ref{eq:ch_ehpc}), per-stage management results in lower total core-hour usage iff the accumulated sum of cores needed at each stage$_i$ is lower than $n$, or: $\sum_{i=1}^{s}{n_i} < n$.
It follows that any workflow with one or more sequential stages and at least one parallel stage can have optimal core-hour usage if per-stage management is used \cite{ehpc}.
%Additionally, in case of asynchronous parallel tasks, E-HPC is able to also grow or shrink resources at run-time. 
%\systemname manages elasticity during workflow execution by dynamically resizing the allocated resources using checkpoint and restart.

Although per-stage management provides lower resource usage, it may negatively affects the total turnaround time (also known as the \textit{makespan}). 
Workflow turnaround time ($T$) can be defined as $T = t + q$, where $t$ is the workflow execution time, and $q$ is the queue waiting time. 
Because resource allocation is performed for each stage in per-stage management, the makespan can be estimated as $T_{Per-Stage} = \sum_{i=1}^{s}{(t_i + q'_i)}$
, where $t_i$ and $q'_i$ are respectively the execution times and queue waiting times of the $i$-th stage.
With a $Big Job$ scheduling strategy, and assuming $t = \sum_{i=1}^{s}{t_i}$, the workflows' makespan is estimated similarly to $T$ above.
Thus, for a per-state management to have lower makespans, the accumulated sum of its waiting times $q'_i$ has to be lower than the single waiting time $q_1$ in the $Big Job$ strategy, i.e., $\sum_{i=1}^{s}{q'_i} < q_1$.
%In other words, per-staged management achieves better turnaround times if the workflow stages have less waiting times than larger big jobs.
One strategy to achieve this is to heuristically pack multiple stages within medium-sized job submissions \cite{zhang2009batch}, though it may not achieve optimal resource usage.
Finally, as the queue waiting time is a system parameter controlled by the resource manager, another natural strategy for the users is to observe its behaviour and estimate it.

\subsection{Challenge: Waiting Time Estimation}

The clear tradeoffs analyzed in the previous subsection show that, in one hand, submitting a large job for execution may have a long single waiting time, with the potential side effect of idle resources during sequential stages.
At the expense of inefficient resource usage, these two characteristics achieve the minimum application runtime possible.
On the other hand, submitting many pilot jobs separately (composing each stage) has the advantage of efficient resource usage as it uses per-staged allocations (as done in E-HPC~\cite{ehpc}).
However, the extra inter-stage waiting times increase the workflow makespan, specially if it interweaves many stages with different resource requirements.
%
%A way to mitigate the increase in workflow makespan is by estimating the queue waiting time.
A way to mitigate this would be to estimate the queue waiting time, with 
a pro-active submission strategy that uses such estimations for coming stages, requesting needed resources during the execution of ongoing stages, 
thus resulting in a minimization of the accumulated inter-stage waiting times \cite{zhang2009batch}. 
However, depending on the estimation accuracy, three outcomes are possible: (i) perfect estimation, (ii) over-estimation, and (iii) under-estimation.
In (i), resource usage and workflow makespan would be optimal.
In (ii), resource usage would be optimal, but a probable increase in workflow makespan would be seen (though less than achieved by per-stage allocation).
In (iii), resources would be ready for use before they are actually needed, and depending on the policy used to mitigate the extra costs regarding this, both the resource usage and workflow makespan would increase.
Some resource managers (such as Slurm \cite{slurm02}) allow job dependency constraints to be specified, including when a job may start execution in case previous jobs have not finished execution.
\section{ASA: The Adaptive Scheduling Architecture}\label{asa:architecture}
In this section we describe the proposed architecture offering a global and unified view of resources to the application, and the proposed algorithm being used to estimate user's queue waiting time for upcoming workflow stages.
\subsection{Architecture}

\begin{figure}
\centering
\includegraphics[width=\columnwidth]{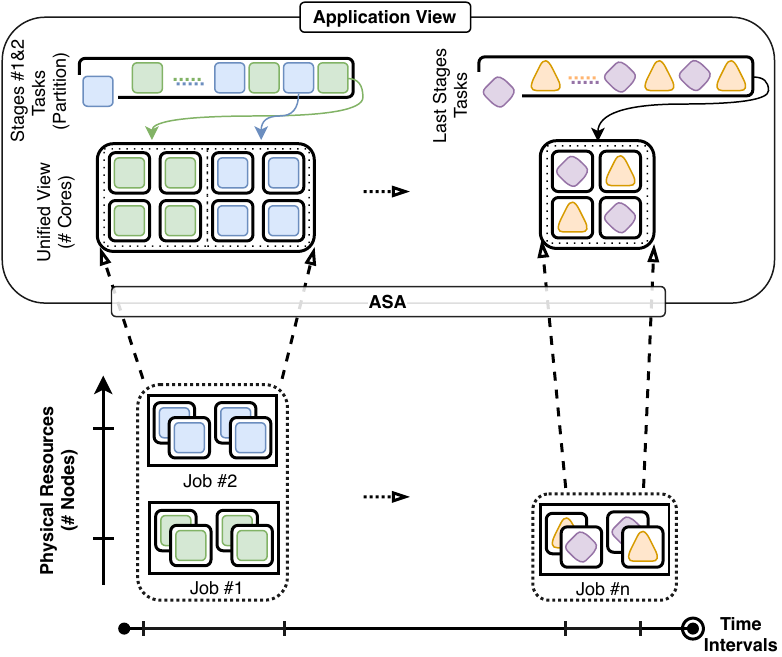}
\caption{ASA - Architecture managing the physical resources. Tasks (the different shapes in the partitions) from different jobs can access resources from multiple jobs. The unified view layer enables users to apply different scheduling strategies, such as pro-active job submissions.}
\label{asa:arch}
\end{figure}

%\comment{My only real criticism is the lack of more detail about how they integrated Mesos with SLURM.}
Figure \ref{asa:arch} illustrates the unified view presented to applications.
Within ASA, 
the \textit{Unified View} layer bridges the management of the physical resources made available through a low level resource manager like Slurm.
Essentially, the application only sees a global pool of resources, where each one can be used freely according to the application's needs.
By extending upon Mesos \cite{2011mesos}, a distributed resource manager, ASA handles scheduling, fault tolerance, resource isolation and control (among collocated tasks), elasticity, and other user defined policies.
Mesos was chosen due to its simplicity and non-intrusiveness at managing resources, allowing users to pack it as a library which can be dynamically loaded.
Moreover, similar resource managers require administrative capabilities to perform similar features, diminishing their portability and usefulness in restricted environments such as HPC clusters.

Current workflows can be easily managed by, and submitted to Mesos.
This can either be achieved through directly submitting the application through a default Mesos executor, or by extending the WMS's internal APIs, bridging them with Mesos and enabling it to manage all workflow's tasks.
Specific scheduling and placement policies can be realized through a Mesos Framework, which is the implementation of a scheduler tailored specifically to an application (e.g. MPI, Spark, etc). 
Mesos then monitors task states (e.g. {\fontfamily{qcr}\selectfont RUN, COMPLETE, FAIL,} etc.) to handle problems such as task crashes, misbehaviours, and unresponsiveness. 
In each case, frameworks can trigger specific actions, e.g., asking for extra resources, or migrating a failed task to another resource.
This model and its associated runtime system enable applications with enhancements such as fault-tolerance, resource isolation, performance control, and the development of novel scheduling algorithms.
For example, one particular feature of this model could be for users who belong to a same project to dynamically share resources with one another and save on total resource consumption for the project.
This can be extra useful at developing phases of a project, where trials and errors are the norm, and performed for testing and prototyping applications.

\subsection{Algorithm}

%\comment{The description of the approach in Section III.B could be clarified. For instance, I don't think the notion of a "mini-batch" is really clear. What is $n$ in this section (the number of workflow levels)? Algorithm should be detailed better in the text, referring to line numbers. I think this section needs a careful proof-reading/rewrite so that the approach is described much more clearly.}

%\comment{Section III B mentions "fixed queue waiting times" of (1s, 10s, 100s, 1000s) -- where do these come from? Is this just an example? Presumably these are not the case in the production experiments.}

\begin{figure}
\centering
\includegraphics[width=\columnwidth]{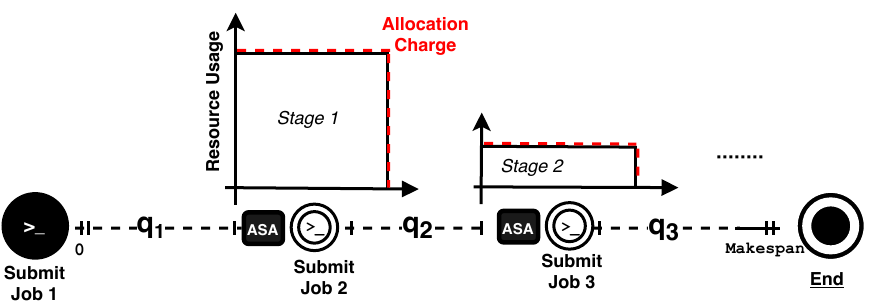}
\caption{ASA - Algorithm workflow illustrating two concurrent pro-active submissions (2 and 3) within ongoing stages. Note the per-staged charging and lower workflow makespan.}
\label{fig:asa_overview}
\end{figure}

Figure \ref{fig:asa_overview} illustrates the algorithm's overall approach.
The algorithm works by maintaining a distribution over a number $m$ of fixed queue waiting times.
For example, for $m=4$, ASA tries to learn which of the four alternatives (indexes) in the vector $(1s, 10s, 100s, 1000s)$ works best as queue waiting time estimation for a given resource allocation request.
Rather than focusing on one specific time such as averaging each perceived queue waiting time, ASA distributes the alternatives following a probability vector $p_t\in[0,1]^m$ with $\sum^m_{t=1} p_t = 1$.
That is, in case one particular alternative works well, for instance index '1s' (or $m_1$), one wants probability $p_t$ to tend to $e_1$ ($p_t\rightarrow e_1 = (1, 0, 0, 0);$ the first unit vector in $R^{m=4}$) when the number of trials $t$ goes to infinity ($t\rightarrow\infty$).
Generally we aim to achieve a good mixture on accuracy and exploration, as a good algorithm needs to be able to detect changes in the queue workload and embed such behaviour in its predictions.
In Reinforcement Learning, the problem of balancing accuracy with discovery is also known as the exploration-exploitation trade off \cite{Thrun92efficientexploration}.

\begin{algorithm}
\caption{ASA - Adaptive Scheduling Algorithm}
\label{alg.asa}
\begin{algorithmic}[1]
\REQUIRE Initialise $\textbf{p}_0$ as $\textbf{p}_{0a} \leftarrow \frac{1}{m}$ for all $m$ actions $a$
\FOR{$t \leftarrow 1,2, \dots$}
\STATE Initialise $\ell_{ta} \leftarrow 0$ for all $a$
\WHILE{$\max_a \ell_{ta} \leq 1$}
\STATE Sample action $a$ according to vector $\textbf{p}_t$
\STATE $\ell_{ta} \leftarrow \ell_{ta} + \ell\left(a\right)$ for action $a$ %(otherwise $\ell_{ta} \leftarrow \ell_{ta}$).
\ENDWHILE
\STATE Update
	\[ \textbf{p}_{t+1,a} \leftarrow e^{- \gamma_t \ell_{ta}} \frac{1}{N_t}\textbf{p}_{t,a}\]
	for all actions $a$. $N_t$ is a normalising factor so that $\sum_a \textbf{p}_a = 1$
\ENDFOR
\end{algorithmic}
\end{algorithm}

%\begin{figure*}\label{asa:results}
%\centering
%\includegraphics[width=0.50\textwidth]{images/montage/MontageMakespan_HPC2N_2.png}
%\includegraphics[width=0.49\textwidth]{images/montage/MontageMakespanAnalysis_UPPMax.png}
%\includegraphics[width=0.49\textwidth]{images/blast/BLAST_MakespanAnalysis_HPC2N.png}
%\includegraphics[width=0.49\textwidth]{images/blast/BLAST_MakespanAnalysis_UPPMax.pdf}
%\includegraphics[width=0.49\textwidth]{images/statistics/MakespanAnalysis_Statistics_HPC2N.pdf}
%\includegraphics[width=0.49\textwidth]{images/statistics/MakespanAnalysis_Statistics_UPPMax.pdf}
%\caption{Montage (HPC2N) Results for 56, 112, and 224 cores - Total Turnaround Analysis (s), Core-hour usage (h), Makespan (s), and Queue Waiting time trends (s).}
%\end{figure*}

This methodology is applied in the following way. 
For each stage at iteration $t$, a waiting time $a$ is estimated for a workflow stage$_y$ and used to submit job$^y$ - the request for change of resources - at time $t_{y-1}-a$,
where $t_{y-1}$ is the expected end-date/deadline of an ongoing workflow stage$_{y-1}$ (Figure \ref{fig:asa_overview}).
If all goes as planned, it is expected that this pro-active job submission strategy minimizes the perceived waiting times between all workflow stages (see Section \ref{sec:background}).
However, if a workflow stage ends later, or an allocation gets assigned earlier than expected, resources may idle for some time before they can be effectively utilized.
Conversely, if a workflow stage ends sooner, or resources become available later than expected, the total workflow process may take longer to complete.
Both can be expressed in terms of a loss function $\ell_{y}(a)$, associated to the workflow stage$_y$ based on the waiting time $a$.
This process is detailed in Algorithm \ref{alg.asa} as follows.

Assume there are $m$ potentially good actions $\{a\}$, as for example, $m$ different time estimations for the queue waiting time. 
Assume also that after each application of an action $a$ to a given case, one can score its loss $\ell(a)$.
The ASA (Adaptive Scheduling Algorithm) consists basically of a double loop.
The outer loop (line 1) iterates over mini batches of cases, referred to as \textit{rounds} and collects as many cases in that round so that the total accumulated loss is bounded.
The inner loop (line 3) iterates over such rounds and ensures that the vector $\ell_t\in \mathbb{R}^m$ collecting loss of the various actions $\{a\}$ is initialised properly before starting a new round, 
and that the vector $\textbf{p}$ is updated properly after each round (line 7).
$e^{\gamma_t}$ is used as a non-increasing sequence, and guarantees the proven convergence of ASA (see Appendix A for more details and the mathematical proof).
$\textbf{p}$ is a distribution over all possible actions $a$ that can be taken: after a while (when it has learned well), it peaks on the best action $a_\ast$, while in the beginning it is spread evenly over all actions $a$.
In other words, in the beginning the algorithm {\em explores} options, like trying out random queue waiting times, or using the resource manager estimate features, while an {\em exploitative} stance is taken when enough evidence is collected, and more accurate estimations can be done.
By following upon these principles, the coming section evaluates how all these concepts link together.
\section{EVALUATION} \label{sec:results}

%In this section, we evaluate ASA and the proposed pro-active scheduling strategy with respect to workflows' total runtime, total resource usage, total queue waiting times, and makespan, which are defined as follows.
In this section, we evaluate ASA's strategy with respect to workflows' total runtime, resource usage, queue waiting times, and makespan, all defined in the following subsections.
We additionally evaluate Algorithm \ref{alg.asa} convergence over time for three different estimation policies for a simulated scenario where the queue waiting time being experienced by the user changes at 5 different points in time (see Subsection~\ref{subsec:convergence}).
At the end of this section, we evaluate ASA estimation accuracy and how it influences the perceived waiting time experienced by applications at runtime.

\subsection{Metrics}

The total runtime is measured by summing up the execution time of each workflow stage.
The summation of each workflow's stage runtime multiplied by the amount of resources used in such stages, measures the total resource usage, or core-hour (measured in hours).
We compare ASA with two different scheduling strategies: (i) traditional, Big Job allocation strategy; and (ii) dynamic, Per-stage job allocation.
As explained in Section~\ref{sec:background}, the Big Job strategy (i) allocates the maximum capacity needed for the entire duration of the workflow, regardless of its stages' needs.
The second strategy (ii), though, allocates resources to workflows in a per-stage manner, for the exact duration of each stage.
Our proposed strategy ASA (iii) pro-actively submits resource changes for a coming stage during the execution of an ongoing stage 
(see Section~\ref{asa:arch}).
The total queue waiting times is thus calculated slightly different in each strategy: in (i), there will be only one queue waiting time (the first one), whereas strategy (ii) has one or more queue waiting times (one additional wait for each workflow stage).
In ASA (iii), the waiting times are measured by the perceived queue waiting times (PWT), i.e., the time interval a coming workflow stage actually waited for resources after a previous stage finished (see Figure~\ref{fig:asa_overview}). 
As this waiting time overlaps with a previous stage execution, the perceived queue waiting time is potentially reduced and can be observed through the makespan metric.
On the other hand, if the perceived queue waiting time is lower than expected, an extra corehour overhead (OH) loss might be incurred.
Then, for each strategy, the total queue waiting time is calculated as the summation of all queue waiting times .
Finally, the total makespan is calculated by subtracting the time the workflow is submitted for execution from the time the workflow successfully finishes execution.
The total makespan takes into consideration all the inter-stage waiting times in each strategy.
We evaluate these metrics for each strategy (Big Job allocation, Per-stage allocations, and ASA) by submitting three different scientific workflows, each with different resource usage and requirement profiles (see Subsection~\ref{workflows}).

\subsection{Computing Systems}

%In order to demonstrate and compare ASA's feasibility, adaptability, and generality features, we run a set of workflows in two different supercomputer centers, with different resource scaling factors: at UPPMax\footnote{https://www.uppmax.uu.se} Rackham, and at HPC2N\footnote{https://www.hpc2n.umu.se} Kebnekaise clusters.
%Rackham comprises 486 nodes with two 10-cores Intel Xeon E5 CPU (v4), with 128 GB memory each.
%Rackham's storage uses the Lustre file system and provides 6.6 PB of storage.
%The interconnect is Infiniband FDR, supportting a theoretical bandwidth of 56 Gb/s and a latency of 0.7 \textit{ms}.
%All Rackham's nodes run CentOS 7.
%Kebnekaise compromises 602 nodes with two 14-cores Intel Xeon E5 CPU (v4), with 128 GB of memory each, and similar Infiniband interconnection as Rackham's.
%Kebnekaise's storage also uses the Lustre file system, providing 2 PB of storage.
%All Kebnekaise's nodes run Ubuntu Xenial (16.04 LTS).

In order to demonstrate and compare ASA's feasibility, adaptability, and generality features, we run a set of workflows in two different supercomputer centers, with different resource scaling factors: at \systemone and \systemtwo.
\systemtwo comprises 486 nodes with two 10-cores Intel Xeon E5 CPU (v4), with 128 GB memory each.
\systemtwo's storage uses the Lustre file system and provides 6.6 PB of storage.
The interconnect is Infiniband FDR, supporting a theoretical bandwidth of 56 Gb/s and a latency of 0.7 \textit{ms}.
All \systemtwo's nodes run CentOS 7, with Slurm 19.05 with its default fair-share scheduling policy.

\systemone compromises 602 nodes with two 14-cores Intel Xeon E5 CPU (v4), with 128 GB of memory each, and similar Infiniband interconnection as \systemtwo's.
\systemone's storage also uses the Lustre file system, providing 2 PB of storage.
All \systemone's nodes run Ubuntu Xenial (16.04 LTS), with Slurm 18.08 with its default fair-share scheduling policy.

\subsection{Applications} \label{workflows}

Three different, real scientific workflows were selected for comparing ASA to Big Job and Per-Stage scheduling strategies, as explained earlier in this section: Montage, BLAST, and Statistics.
\\
\\
\noindent\textbf{Montage} \cite{montage} is a data intensive application that constructs the mosaic of a sky survey. 
The workflow has nine ordered stages, grouped into two parallel (first two, and fifth) and two sequential (third and fourth, and last three) stages (Figure~\ref{fig:montage}).
All runs of Montage construct an image for survey {\fontfamily{qcr}\selectfont M17} on {\fontfamily{qcr}\selectfont band j}, {\fontfamily{qcr}\selectfont degree 8.0} from the {\fontfamily{qcr}\selectfont 2mass} Atlas images.
\\
\\
\noindent\textbf{BLAST} \cite{altschul1997gapped} is a compute intensive applications that matches DNA sequences against a large ( $>$ 6 GB) sequence database. 
The workflow splits an input file (of few KBs) into several smaller files and then uses parallel tasks to compare the input against the large sequence database. 
BLAST is composed of two main stages: one parallel and one sequential.
The database is loaded in-memory on all  compute nodes during the parallel stage. 
Finally, all the outputs from the parallel stage are merged into a single file (sequential stage).
\\
\\
\noindent \textbf{Statistics} \cite{kolter2011redd} is an I/O and network intensive application that calculates various statistical metrics (mean, median, average, standard deviation, variance, etc.) from a large dataset with measurements of electric power consumption in a household with an one-minute sampling rate over a period of almost 4 years. 
Different electrical quantities and some sub-metering values are available in a public dataset.
The statistics workflow is composed mainly of a two sequential and two parallel stages, intertwined, consuming most of the processing due to communication among the parallel tasks.
\\
\\
\noindent \textbf{Workflow configuration}.
For each one of the three strategies (Big Job, Per-stage, and ASA), these three workflows are submitted sequentially to the queue, concurrently one after the other. This was done using six different scaling factors: In \systemone, workflows use 28, 56, and 112 cores, respectively; whereas in \systemtwo, workflows use 160, 320, and 640 cores, respectively.
This combination creates a total of 54 different runs.
For ASA's strategy, Algorithm \ref{alg.asa}'s state is kept across different runs, meaning % the probability distribution $p$, the accumulated losses $r$, the current loss (regarding a job submission) $\ell_y$, and the time intervals $m$ are 
all of its variables are shared among the different workflow submissions.
This allows the algorithm to converge and adapt itself more quickly to the current queue state, minimizing errors.
Finally, in this evaluation the loss function $\ell_{y}(a)$ for a given job geometry $y$ is defined as \begin{align}
\label{asa_loss}
   \ell_{y}(a):= \begin{cases} 0, & optimal; \\ 1, & otherwise \end{cases}
\end{align} \noindent where \textit{optimal} means the algorithm sampled the best possible (closest to the true queue waiting time) action $a$ among the $m$ alternatives available, and thus it returns a loss of $0$, and $1$ otherwise.
Although more complex functions could be used, choosing a simple loss function allows ASA's behaviour to be understood more easily.
Moreover, queue waiting times can be very large at some supercomputer centers, as it depends on resource availability and on many job constraints.
As mentioned in Algorithm \ref{alg.asa}, lengthier $m$'s should theoretically return more accurate estimations. 
However, for the purposes of this evaluation and due to practical runtime reasons, $m$ is empirically set to represent a maximum queue waiting time of ${\sim}$28 hours (100k seconds), since this was the maximum queue waiting time reported in Systems 1 and 2.
Thus, the value of $m = 53$ is used in Algorithm \ref{alg.asa} to split the possible range estimators in 53 time intervals representing possible queue waiting time alternatives. 
The alternatives cover multiples of 10's, 100's, 1k's, 10k's, and 100k time intervals (in seconds), with higher number of alternatives assigned to values 10's and 100's due to the higher queue waiting times variability usually faced by smaller jobs (with up to 112 assigned cores) usually falling down in these ranges.
%Finally, all three workflows use the Tigres\footnote{http://tigres.lbl.gov} WMS for runtime execution.
Finally, all three workflows use the Tigres WMS\footnote{http://tigres.lbl.gov} for runtime execution, and the Per-Stage submissions use Tigres' E-HPC feature.

\begin{figure}[t]
\centering
\includegraphics[width=\linewidth]{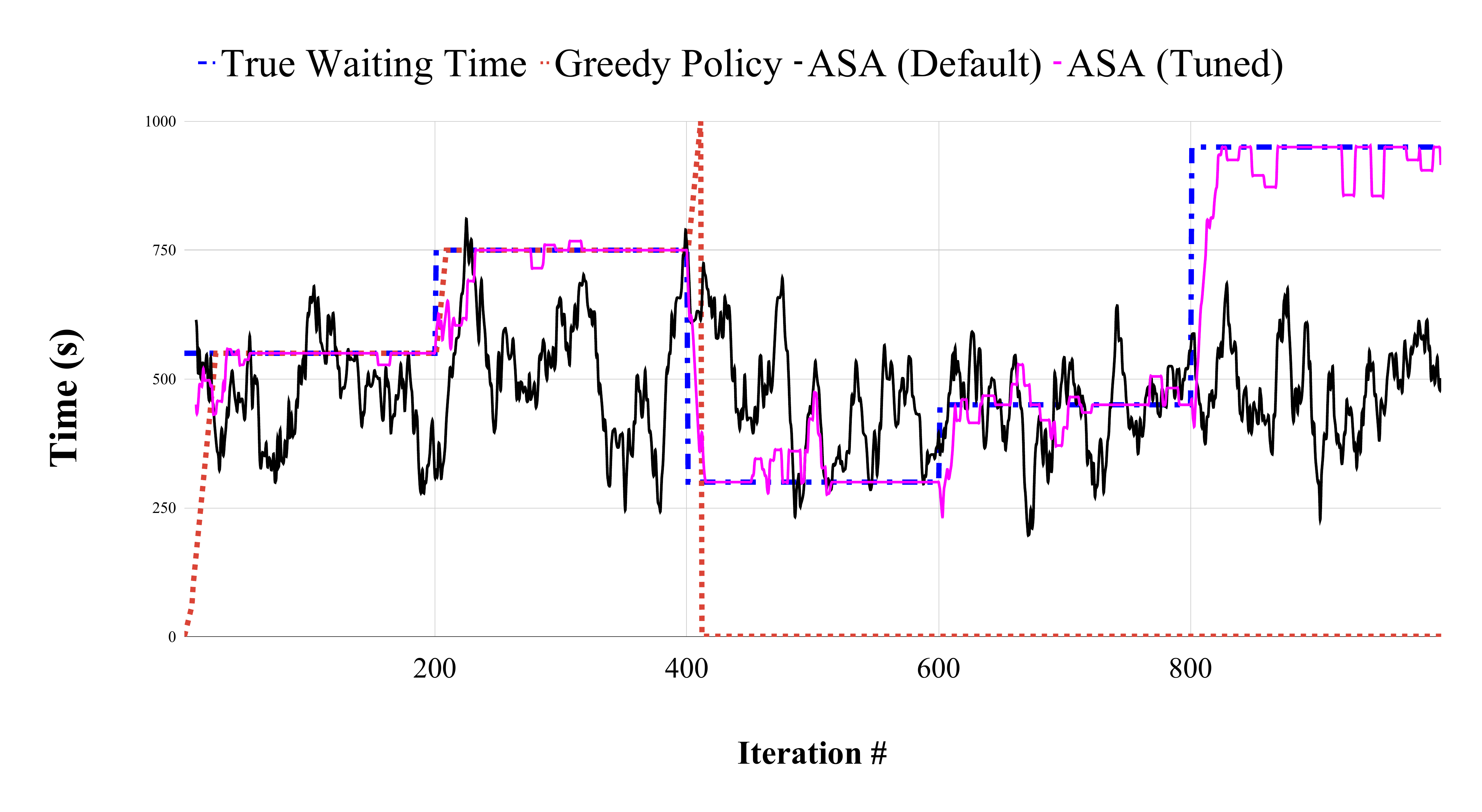}
\caption{ASA's estimation convergence over time regarding queue waiting time (dark dashed blue line) with three different sampling policies: Greedy (red dotted line), ASA's default (black line), and ASA tuned (light pink line).
}
\label{asa:convergence}
\end{figure}

\subsection{Convergence Results}\label{subsec:convergence}

Figure~\ref{asa:convergence} shows a 1000 iterations simulation demonstrating how ASA 
(Algorithm~\ref{alg.asa}) waiting time estimations converges in a hypothetical scenario where the true waiting time changes over time.
To test ASA's adaptability capabilities, the true waiting time (blue stepped line) is randomly varied at five different occasions: at iterations numbers 0, 200, 400, 600, and 800.
The default ASA policy (black thick line) takes rather too many iterations to converge to the true waiting time, which suddenly changes and worsens its convergence trend.
It does so because it keeps exploring the interval space in order to validate its knowledge.
However, with a tuned policy (pink thin line), where the perceived queue waiting times are used to randomly and repeatedly adjust the probability distribution $p$ (used in Algorithm \ref{alg.asa}) with the calculated losses, the convergence velocity changes drastically.
As it can be seen, at every true waiting time variation the tuned policy strategy enables ASA to converge to the true waiting time more rapidly.
Even though, it still allows ASA to keep exploring the interval space, though it makes fewer miss predictions than the default sampling policy.
A greedy approach is also shown (dashed red line), where the minimum perceived loss is always used for making estimations for the waiting times.
Because the simple loss function $\ell(i)$ (see definition (\ref{asa_loss}) in Section \ref{asa:architecture}) is used, when the true waiting time suddenly drops, the greedy policy reaches a local minimum, and does not behave correctly afterwards, defaulting to a very conservative loss estimator (i.e., every pro-active submission happens at the end of a stage, similarly to the Per-Stage's strategy) and thus behaving as if the algorithm was not used at all.

\subsection{Sensitivity Analysis}

By default, ASA uses the resource manager's helpers to set job dependencies.
This allows ASA to specify different dependencies between sorted workflow stages with the advantage of optimizing resource usage, as a job cannot have its resources allocated (and thus charged) until all dependent jobs completed execution.
However, this can also cause ASA to deviate towards non-optimal estimations, and furthermore randomly defer the start of jobs.
Thus, to illustrate how ASA behaves in environments managed by resource managers with no support of job dependency helpers, and to calculate such impacts on the total workflow resource usage, an experiment without this setting is evaluated in \systemone for the Montage workflow with 112 cores. 
This strategy is henceforce denoted \emph{ASA Na\"ive}.

%\comment{Unfortunately it is hard to tease out from the results (and the description thereof) how accurate the queue wait time predictions are. We're essentially given makespan results. Section IV.B defines a "loss function", which seems like a good way to quantify the answer to the "how good was the prediction?" question. But then this metric is not discussed at all in the result.}
%\comment{The text does not mention whether the results are averages over multiple trials. This is obviously a huge concern.}

\begin{figure*}
    \centering
    \subfigure[\systemone]
    {
        \includegraphics[width=0.48\linewidth]{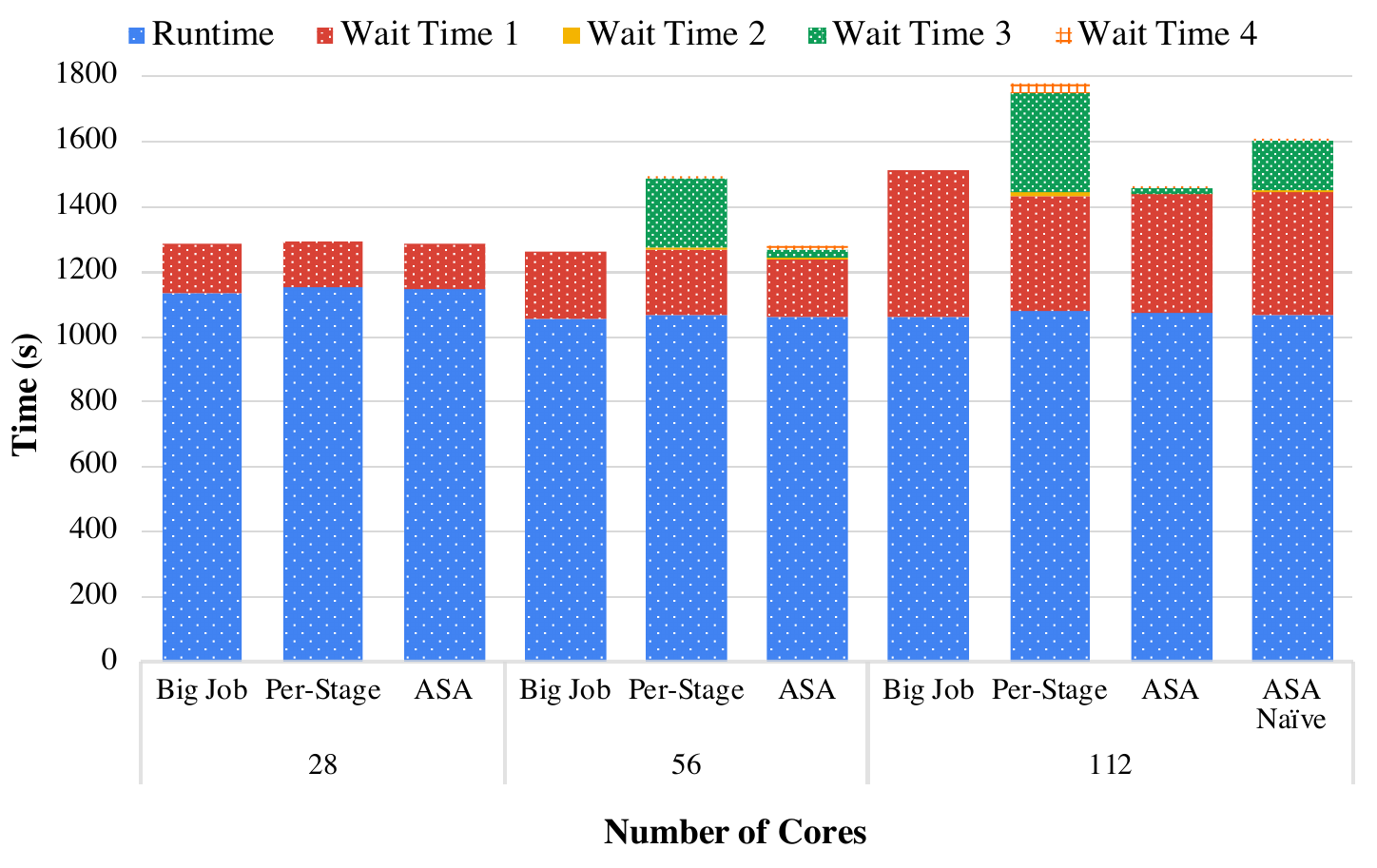}
        \label{fig:montage_hpc2n}
    }
    \subfigure[\systemtwo]
    {
        \includegraphics[width=0.48\linewidth]{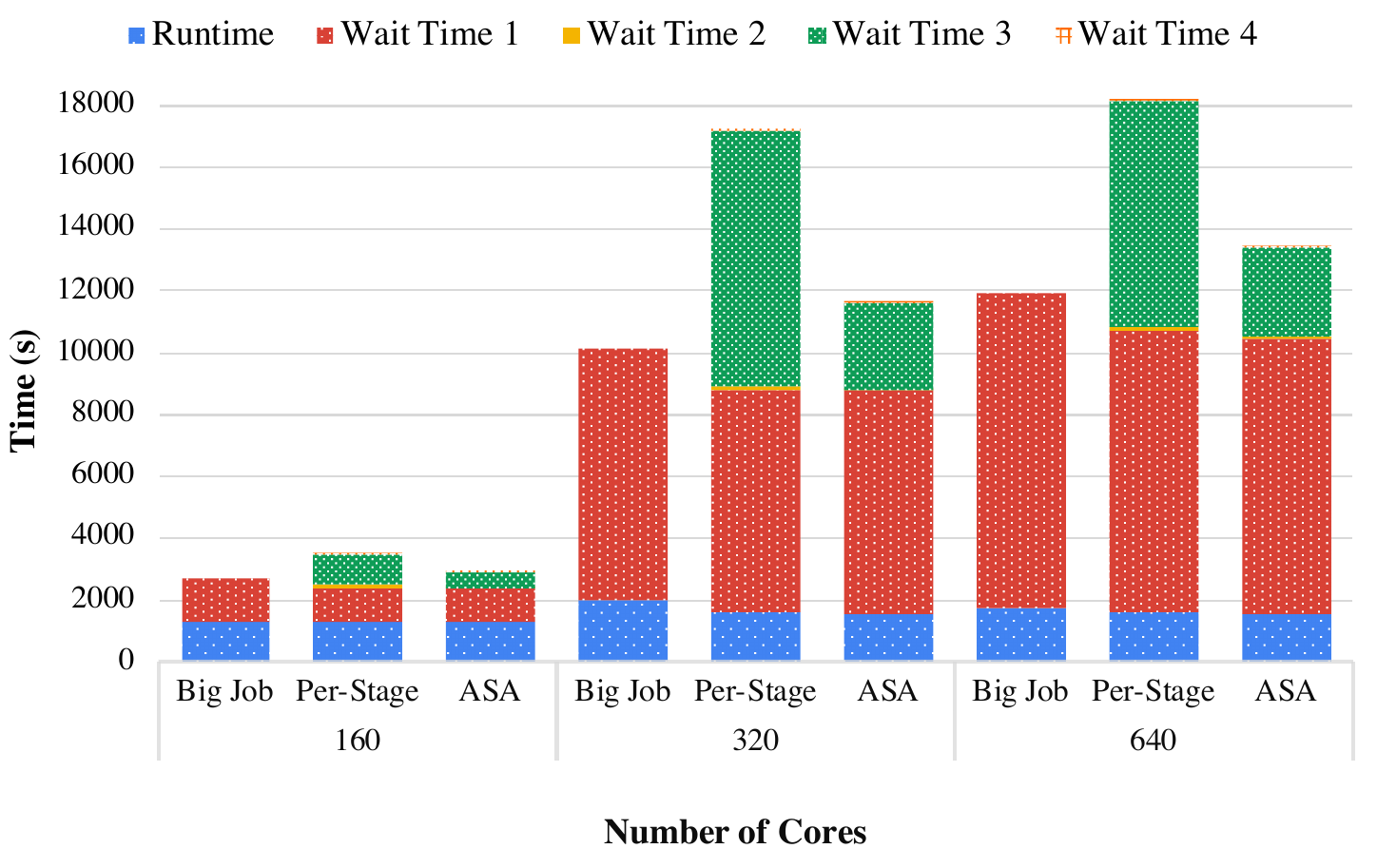}
        \label{fig:montage_uppmax}
    }
    \caption{Montage Workflow - Makespan results for (a) \systemone and (b) \systemtwo for different scaling factors (28, 56, 112, 160, 320, and 640 cores), and scheduling strategies (Big Job, Per-Stage, and ASA). Number of cores indicate peak allocations for a given strategy. ASA Naive strategy means no resource manager dependency setting is used.}
    \label{fig:montage_makespan}
\end{figure*}

\begin{figure*}
    \centering
    \subfigure[\systemone]
    {
        \includegraphics[width=0.48\linewidth]{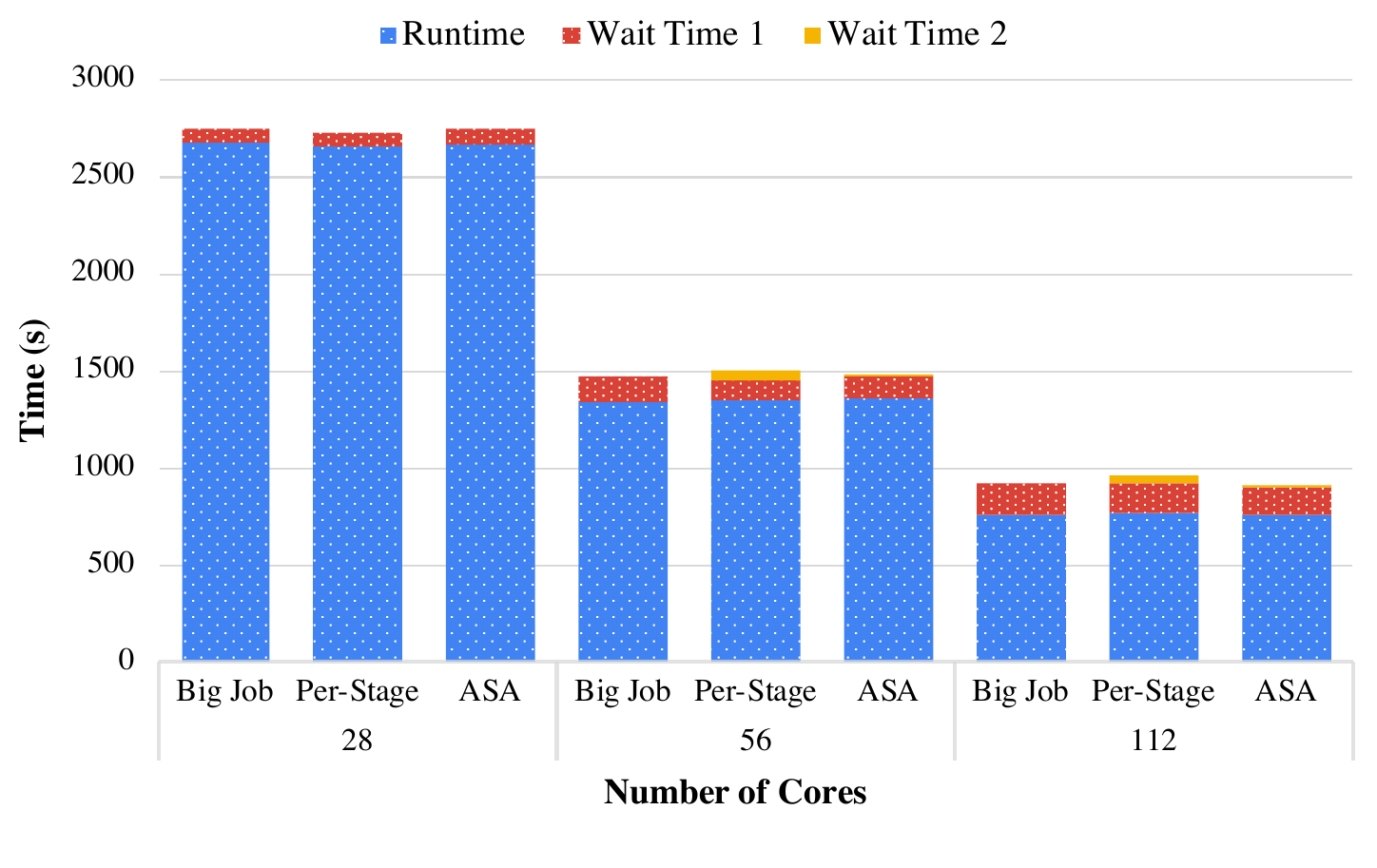}
        \label{fig:blast_hpc2n}
    }
    \subfigure[\systemtwo]
    {
        \includegraphics[width=0.48\linewidth]{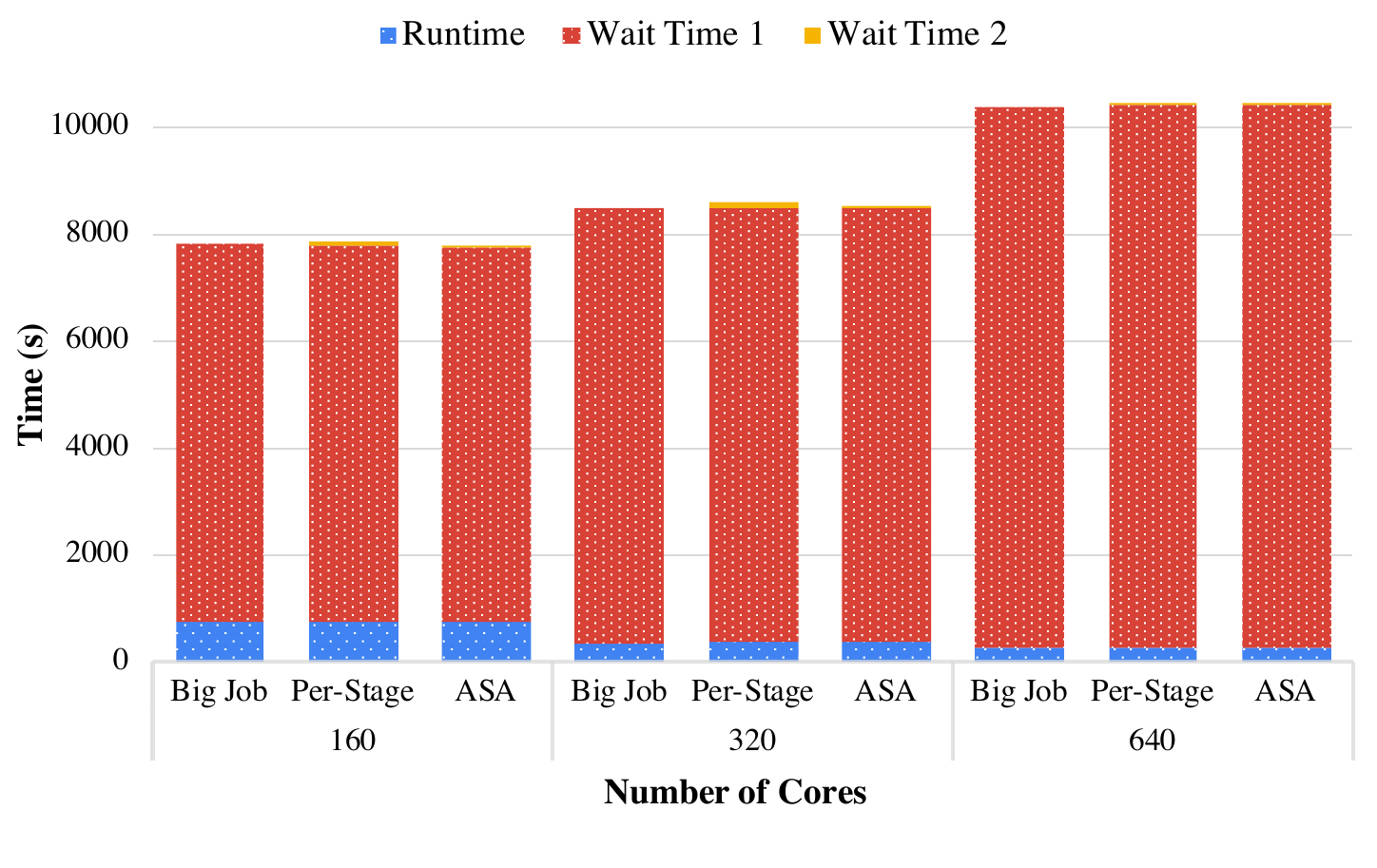}
        \label{fig:blast_uppmax}
    }
    \caption{BLAST Workflow - Makespan results for (a) \systemone and (b) \systemtwo for different scaling factors (28, 56, 112, 160, 320, and 640 cores), and scheduling strategies (Big Job, Per-Stage, and ASA). Number of cores indicate peak allocations for a given strategy.}
    \label{fig:blast_makespan}
\end{figure*}

\begin{figure*}
    \centering
    \subfigure[\systemone]
    {
        \includegraphics[width=0.48\linewidth]{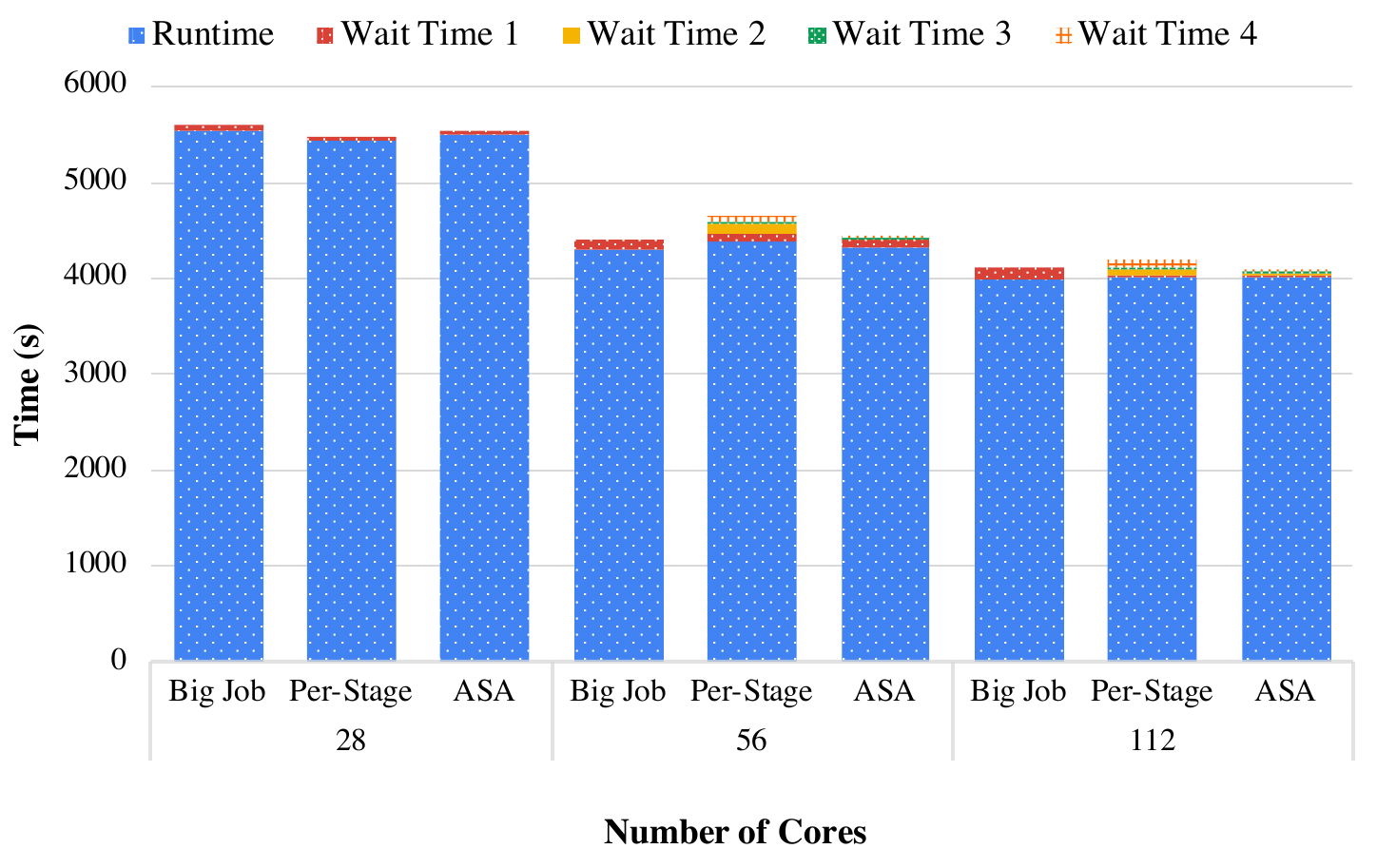}
        \label{fig:stats_hpc2n}
    }
    \subfigure[\systemtwo]
    {
        \includegraphics[width=0.48\linewidth]{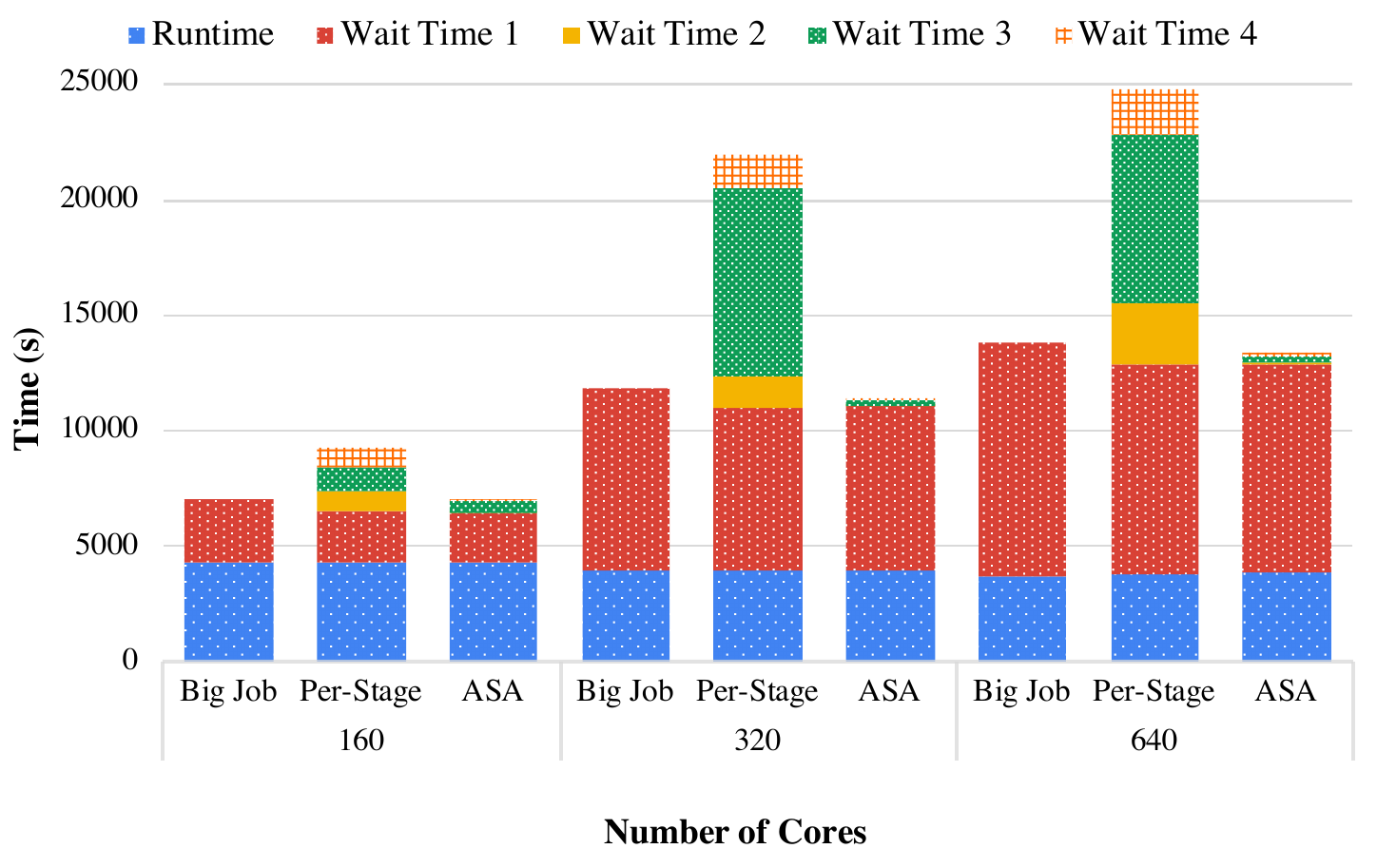}
        \label{fig:stats_uppmax}
    }
    \caption{Statistics Workflow - Makespan results for (a) \systemone and (b) \systemtwo for different scaling factors (28, 56, 112, 160, 320, and 640 cores), and scheduling strategies (Big Job, Per-Stage, and ASA). Number of cores indicate peak allocations for a given strategy.}
    \label{fig:stats_makespan}
\end{figure*}

A large repetition number within the ASA tuned sampling policy has the effect of influencing (or biasing) ASA to follow the last observed waiting time, and thus this feature should be used with caution to not make ASA simply follow (or exploit) its first queue waiting time observations, devoting it of learning new outcomes and changes in the queue workload.
In the following sub-sections, Algorithm \ref{alg.asa} is tuned with a repetition parameter of 50, same value used for the previous simulation shown in Figure \ref{asa:convergence}.
%\devarshi{where is this figure mentioned in the text?}

As a complement to these empirical results, we proof the theoretical convergence of the ASA algorithm in Appendix~A. 

\subsection{Makespan Results}

Figures \ref{fig:montage_makespan}, \ref{fig:blast_makespan}, and \ref{fig:stats_makespan} show all three workflows' makespan breakdowns, showing the different inter-stage queue waiting times for each scheduling strategy (Big Job, Per-Stage, and ASA allocations), and for the six different number of cores (scaling, representing the peak allocations for each strategy).
The figures are split in two columns, (a) left and (b) right, which show results for \systemone and \systemtwo, respectively.
The difference in the queue waiting times between the two supercomputer centers are substantial, with higher waiting times in \systemtwo.

Figures \ref{fig:montage_makespan}(a) and (b) show the makespan for Montage.
It is noticeable that early stages suffer from higher queue waiting times due to Montage beginning execution with a parallel stage.
The total waiting time gets worse as the peak core allocation (Number of cores) scales, negatively impacting the Per-Stage strategy in every scenario with more than 28 cores.
Due to multiple interactions of concurrently running workflows, dynamically updating how the queue is behaving ($p$ distribution in Algorithm \ref{alg.asa}) are required. 
ASA's strategy leverages this information to submit resource changes earlier, lowering the total waiting times.
This is noticeable already at scaling to 56 and 112 cores, where the total makespan is lower than for the Big Job allocation.
For Montage Naive, with 112 cores at \systemone, there were additional delays as resources for the third stage got ready for use before the second stage had completed.
In this case, ASA canceled the submission and re-submitted the job for the third stage, which incurred in an additional (perceived) queue waiting time as it can be noticed in the larger \textit{Wait Time 3} in Figure \ref{fig:montage_hpc2n}.

However, the same effect does not happen to BLAST, as can be seen in figures \ref{fig:blast_makespan} (a) and (b).
As BLAST is a two stages workflow, where the first parallel stage is considerable larger than the second sequential one, the effects of using different scheduling strategies are neglectable.
Besides that, BLAST is a very scalable application, essentially keeping the resource utilization high as the core scaling factor increases. Furthermore, due to the higher queue waiting times at \systemtwo, BLAST's total makespan gets severely impacted.
Finally, more notable dynamics are perceived in the Statistics workflow (Figures \ref{fig:stats_makespan} (a) and (b)).
This is a four stage and network intensive workflow, with large execution times in each stage.
At first, in \systemone (Figure \ref{fig:stats_hpc2n}), the queue waiting time has limited impact on the total makespan.
However, at the busier center (\systemtwo, Figure \ref{fig:stats_uppmax}) with larger queue waiting times, the learnt information gathered by other concurrent workflows allows ASA's pro-active system to essentially submit future resource change jobs earlier than even upcoming stages. This has significant effect, specially at 320 and 640 cores, where the queue waiting times are severely impacted for the Per-Stage strategy, which sometimes had twice the makespan of Big Job allocations.

\subsection{Resource Usage and ASA Performance}

Table \ref{tab:results} summarizes all runs and the measured metrics explained in previous sections.
Below each workflow, a normalized average of collected metrics shows workflow's results in overview.
This average is related to the lowest metric for each resource scaling row. Besides that, percentages inside parentheses represent the extra times incurred when comparing specific metrics with the best metric for that resource scaling.

It is visible that Per-Stage and ASA strategies provide the best resource usage in most scenarios, as shown in the table as Core-hour Usage. This comes, however, at the cost of extra workflow's makespan.
Also, due to some variations in how long the workflows take to run, the resource usage in each strategy can vary.
As noted in the previous subsection, the total makespan can be severely impacted when the amount of time an application waits in the queue is larger than the application's total execution time.
As Montage is a not a scalable application, its execution time across different scaling factors does not considerably decrease its total execution time, and requesting larger amounts of resources actually impacts the total makespan negatively.
This is illustrated by the Montage and BLAST workflows (320 and 640 cores), where the Per-Stage allocation strategy had 82\% and 13\% increase in the makespan. 
Pro-active ASA submissions reduced these severe extra times by 72\% and 9\% respectively, reducing large queue waiting time impacts as the normalized average makespans show.
This behavior is particularly more noticeable for runs using the Per-Stage strategy than it is for ASA, as it learns about the queue's current state by observing the impacts other concurrent workflow submissions had.
This information allows ASA to act to mitigate such severe impacts in earlier stages.

% Please add the following required packages to your document preamble:
% \usepackage{booktabs}
% \usepackage{multirow}
% \usepackage{graphicx}
\begin{table*}
\caption{Experimental results respectively for Montage, BLAST, and Statistic Workflows in the six different core scalings. Bold values show best normalized results in relation to Big Job, Per-Stage, and ASA allocations. Normalized averages shown below results for each workflow -- the lower, the better --, with extra percentages (in relation to the \emph{best} result achieved for any strategy) included inside parentheses (values over 1\%).\\Acronyms: WF (Workflow), TWT (Total Waiting Time), and CH (Core-Hour).}
\label{tab:results}
\centering
\resizebox{\textwidth}{!}{%
\begin{tabular}{@{}r|r|rrr|rrr|rrr@{}}
\toprule
\multicolumn{2}{c|}{} & \multicolumn{3}{c|}{\textbf{Big Job Allocation}} & \multicolumn{3}{c|}{\textbf{Per-Stage Allocation}} & \multicolumn{3}{c}{\textbf{ASA}} \\ \midrule
%\multicolumn{1}{c|}{\textbf{WF}} & \textbf{Scaling} & \multicolumn{1}{c|}{\textbf{\begin{tabular}[c]{@{}c@{}}Waiting\\ Time (s)\end{tabular}}} & \multicolumn{1}{r|}{\textbf{\begin{tabular}[c]{@{}r@{}}Makespan \\ (s)\end{tabular}}} & \textbf{Core-hour (h)} & \multicolumn{1}{c|}{\textbf{\begin{tabular}[c]{@{}c@{}}Waiting\\ Time (s)\end{tabular}}} & \multicolumn{1}{c|}{\textbf{Makespan (s)}} & \textbf{Core-hour (h)} & \multicolumn{1}{c|}{\textbf{\begin{tabular}[c]{@{}c@{}}Waiting\\ Time (s)\end{tabular}}} & \multicolumn{1}{c|}{\textbf{Makespan (s)}} & \textbf{Core-hour (h)} \\ \midrule
\multicolumn{1}{c|}{\textbf{WF}} & \textbf{Cores} & \textbf{\begin{tabular}[c]{@{}r@{}}TWT (s)\end{tabular}} & \textbf{Makespan (s)} & \textbf{\begin{tabular}[c]{@{}r@{}}CH (h)\end{tabular}} & \textbf{\begin{tabular}[c]{@{}r@{}}TWT (s)\end{tabular}} & \textbf{Makespan (s)} & \textbf{\begin{tabular}[c]{@{}r@{}}CH (h)\end{tabular}} & \textbf{\begin{tabular}[c]{@{}r@{}}TWT (s)\end{tabular}} & \textbf{Makespan (s)} & \textbf{\begin{tabular}[c]{@{}r@{}}CH (h)\end{tabular}} \\ \midrule
\multirow{6}{*}{\rotatebox[origin=c]{90}{\textbf{\underline{Montage}}}} & 28 & 150 (+13\%) & 1287 & 9 (+24\%) & 258 (+95\%) & 1408 (+10\%) & 7 & \textbf{132} & \textbf{1277} & \textbf{7} \\
 & 56 & \textbf{206} & \textbf{1261} & 16 (+32\%) & 426 (+105\%) & 1496 (+19\%) & 12 & 219 (+6\%) & 1280 (+2\%) & \textbf{12} \\
 & 112 & 452 (+15\%) & 1513 (+3\%) & 33 (+65\%) & 699 (+78\%) & 1779 (+22\%) & 20 & \textbf{393} & \textbf{1464} & \textbf{20} \\
 & 160 & \textbf{1415} & \textbf{2718} & 58 (+47\%) & 2220 (57\%) & 3507 (+29\%) & 40 & 1652 (+17\%) & 2921 (+7\%) & \textbf{39} \\
 & 320 & \textbf{8135} & \textbf{10126} & 177 (+68\%) & 15582 (+91\%) & 17170 (+69\%) & 106 & 10062 (+24\%) & 11637 (+12\%) & \textbf{105} \\
 & 640 & \textbf{10200} & \textbf{11940} & 309 (+83\%) & 16600 (+63\%) & 18200 (+52\%) & 171 & 11851 (+16\%) & 13436 (+12\%) & \textbf{169} \\ \midrule
\multicolumn{2}{c|}{\textit{\textbf{\begin{tabular}[c]{@{}c@{}}Normalized \\ Average\end{tabular}}}} & \textit{\textbf{+5\%}} & \textit{\textbf{+1\%}} & \textit{+53\%} & \textit{+82\%} & \textit{+34\%} & \textit{0\%} & \textit{+10\%} & \textit{+6\%} & \textit{\textbf{0\%}} \\ \midrule
\multirow{6}{*}{\rotatebox[origin=c]{90}{\underline{\textbf{BLAST}}}} & 28 & 70 (+3\%) & 2750 & 20 & \textbf{68} & 2727 & \textbf{20} & 75 & 2749 & 20 \\
 & 56 & 133 (+20\%) & \textbf{1476} & 20 & 153 (+2\%) & 1508 & 21 & \textbf{111} & 1477 & \textbf{21} \\
 & 112 & 165 (+15\%) & 926 (+2\%) & 23 & 194 (+35\%) & 965 (+6\%) & 24 & \textbf{144} & \textbf{907} & \textbf{23} \\
 & 160 & 7100 & 7846 & 33 & 7125 & 7880 & 33 & \textbf{7041} & \textbf{7800} & \textbf{33} \\
 & 320 & \textbf{8133} & \textbf{8494} & 32 & 8240 & 8611 & 33 & 8194 & 8557 & \textbf{32} \\
 & 640 & \textbf{10133} & 10394 & \textbf{46} & 10150 & 10429 & 47 & 10144 & 10419 & 48 \\ \midrule
\multicolumn{2}{c|}{\textit{\textbf{\begin{tabular}[c]{@{}c@{}}Normalized \\ Average\end{tabular}}}} & \textit{+6\%} & \textit{\textbf{0\%}} & \textit{+1\%} & \textit{+13\%} & \textit{+2\%} & \textit{\textbf{0\%}} & \textit{\textbf{+2\%}} & \textit{0\%} & \textit{+1\%} \\ \midrule
\multirow{6}{*}{\rotatebox[origin=c]{90}{\textbf{\underline{Statistics}}}} & 28 & 52 (+8\%) & 5593 (+2\%) & 43 (+2\%) & \textbf{48} & \textbf{5487} & \textbf{42} & 51 (+6\%) & 5549 & 43 \\
 & 56 & \textbf{96} & \textbf{4397} & 66 (+99\%) & 263 (+174\%) & 4644 (+6\%) & 34 & 116 (+21\%) & 4444 & \textbf{33} \\
 & 112 & 124 (+77\%) & 4110 (+1\%) & 124 (+99\%) & 191 (+173\%) & 4193 (3\%) & 62 & \textbf{70} & \textbf{4085} & \textbf{62} \\
 & 160 & 2772 (+2\%) & 7095 (+2\%) & 192 (+102\%) & 4960 (+83\%) & 9241 (+32\%) & 95 & \textbf{2712} & \textbf{6986} & \textbf{95} \\
 & 320 & 7935 (+6\%) & 11886 (3\%) & 351 (+99\%) & 18008 (+141\%) & 21993 (+92\%) & 177 & \textbf{7471} & \textbf{11439} & \textbf{176} \\
 & 640 & 10122 (+7\%) & 13868 (+4\%) & 665 (+97\%) & 21014 (+121\%) & 24817 (+86\%) & \textbf{338} & \textbf{9497} & \textbf{13369} & 344 \\ \midrule
\multicolumn{2}{c|}{\textit{\textbf{\begin{tabular}[c]{@{}c@{}}Normalized \\ Average\end{tabular}}}} & \textit{+17\%} & \textit{+2\%} & \textit{+83\%} & \textit{+115\%} & \textit{+36\%} & \textit{\textbf{0\%}} & \textit{\textbf{+5\%}} & \textit{\textbf{0\%}} & \textit{+1\%} \\ \bottomrule
\end{tabular}
}
\end{table*}

\subsection{Prediction accuracy}

In order to validate and quantify ASA's decision accuracy that has impact on its performance as a scheduling algorithm, additional experiments were conducted. 
Here, each job geometry related to each workflow described in Section~\ref{workflows} is submitted to their respective system 60 times, with a one minute time interval between submissions. 
This is done to capture variations in each system's queue workload, which affects the experienced waiting times.
As in the previous evaluations, \systemone handles all job geometries submissions with 28, 56, and 112 cores, whereas \systemtwo handles job geometries with 160, 320, and 640 cores.
For each submission, waiting times are compared to ASA predictions of waiting time. 
Table \ref{tab:summary} summarizes averages results for each workflow job geometry.
In this table, the real waiting time (WT) averages actual queue waiting times (in hours), ASA WT averages predicted waiting times (in hours), and Perceived WT averages workflows' actual waiting times (in hours) are given. 
The impact of the predictions are also assessed as follows: 
Hit (the higher, the better) and Miss (the lower, the better) ratios represent the fractions of ASA's accurate- and over- predictions. 
The latter is increased when jobs need to be re-submitted due to larger predictions than actual WTs, over all job submissions.
Misses and over-predictions impact total resource usage (measured in core-hours) because job allocations get assigned earlier than the estimates, causing extra job submission overheads (OH) when compared to resource usage for the Per-Stage strategy.

As it can be seen, there are high variations in \systemone (Cores 28-112), whereas there are no misses (incorrect predictions causing re-submissions) at all for \systemtwo due to its stability.
This can be explained due to the higher fragmentation caused by smaller jobs and allocations.
Notable, although there is a fair amount of misses for smaller job geometries (up to 112 cores) and thus job re-submissions for \systemone, ASA still controls considerably the core-hour overhead losses.
We remark that ASA achieves very good overall results on \systemone, as summarized in Table~\ref{tab:results}, where ASA has very good makespan and resource usage results compared to the alternative strategies (Big Job and Per-Stage). 

\begin{table}
\caption{ASA - Average results for the Montage, BLAST and Statistic workflows across six job geometries. Cores 28, 56, and 112 are \systemone's, whereas 160, 320, and 640 are \systemtwo's.\\Acronyms: WT (Waiting Time), PWT (Perceived Waiting Time), and OH (Core-Hour overhead in hours).}
\label{tab:summary}
\centering
\resizebox{\linewidth}{!}{%
\begin{tabular}{@{}rr|rrrrrr@{}}
\toprule
%\multicolumn{2}{c|}{} & \multicolumn{6}{c|}{\textbf{Job Geometry 1}} \\ \midrule

\multicolumn{1}{c}{\textbf{}} & \textbf{Cores} & \textbf{\begin{tabular}[c]{@{}r@{}}Real\\WT (h)\end{tabular}} & \textbf{\begin{tabular}[c]{@{}r@{}}ASA\\WT (h)\end{tabular}} & \textbf{\begin{tabular}[c]{@{}r@{}}ASA\\PWT (h)\end{tabular}} & \textbf{\begin{tabular}[c]{@{}r@{}}Hit\\Ratio (\%)\end{tabular}} & \textbf{\begin{tabular}[c]{@{}r@{}}Miss\\Ratio (\%)\end{tabular}} & \textbf{\begin{tabular}[c]{@{}r@{}}OH \\Loss (h)\end{tabular}} \\ \midrule
\multirow{6}{*}{\rotatebox[origin=c]{90}{\textbf{\underline{Montage}}}} & 28 & 0.4$\pm$0.3 & 0.7$\pm$0.6 & 0.5$\pm$0.4 & 60 & 40 & 1.7$\pm$0.5 \\
 & 56 & 1.1$\pm$0.8 & 1.2$\pm$0.9 & 0.4$\pm$0.4 & 68 & 32 & 3.0$\pm$0.8 \\ 
 & 112 & 1.5$\pm$0.7 & 2.0$\pm$1.9 & 0.5$\pm$0.4 & 87 & 13 & 2.0$\pm$0.8 \\
 & 160 & 11 $\pm$1.6 & 3.9$\pm$4.6 & 0.7$\pm$0.3 & 100 & 0 & 0 \\
 & 320 & 15 $\pm$1.3 & 12$\pm$3.9 & 0.2$\pm$0.3 & 100 & 0 & 0 \\
 & 640 & 17 $\pm$0.6 & 12$\pm$3.3 & 0.3$\pm$0.2 & 100 & 0 & 0 \\ \midrule
%\multicolumn{2}{c|}{\textit{\textbf{\begin{tabular}[c]{@{}c@{}}Average\end{tabular}}}} & \textit{\textbf{+5\%}} & \textit{\textbf{+1\%}} & \textit{+53\%} & \textit{+82\%} & \textit{+34\%} & \textit{0\%} \\ \midrule
\multirow{6}{*}{\rotatebox[origin=c]{90}{\underline{\textbf{BLAST}}}} & 28 & 0.4$\pm$0.3 & 1.0$\pm$1.0 & 0.6$\pm$0.3 & 70 & 30 & 8$\pm$1.9 \\
 & 56 & 1.1$\pm$0.8 & 1.3$\pm$1.2 & 0.7$\pm$0.5 & 71 & 29 & 11$\pm$2.7 \\
 & 112 & 1.5$\pm$0.7 & 1.0$\pm$1.0 & 0.6$\pm$0.4 & 89 & 11 & 3$\pm$0.7 \\
 & 160 & 11$\pm$1.6 & 4.5$\pm$5.0 & 0.7$\pm$0.4 & 100 & 0 & 0 \\
 & 320 & 15$\pm$1.3 & 11$\pm$4.1 & 0.2$\pm$0.3 & 100 & 0 & 0 \\
 & 640 & 16$\pm$0.6 & 11$\pm$3.8 & 0.3$\pm$0.2 & 100 & 0 & 0 \\ \midrule
%\multicolumn{2}{c|}{\textit{\textbf{\begin{tabular}[c]{@{}c@{}}Average\end{tabular}}}} & \textit{+6\%} & \textit{\textbf{0\%}} & \textit{+1\%} & \textit{+13\%} & \textit{+2\%} & \textit{\textbf{0\%}} \\ \midrule
\multirow{6}{*}{\rotatebox[origin=c]{90}{\textbf{\underline{Statistics}}}} & 28 & 0.4$\pm$0.6 & 0.5$\pm$0.7 & 0.4$\pm$0.4 & 67 & 33 & 3$\pm$0.2 \\
 & 56 & 1.1$\pm$0.8 & 1.2$\pm$0.9 & 0.4$\pm$0.4 & 69 & 31 & 6$\pm$2.0 \\
 & 112 & 1.5$\pm$0.7 & 2.0$\pm$1.9 & 0.5$\pm$0.4 & 87 & 13 & 5$\pm$1.0 \\
 & 160 & 11$\pm$1.7 & 5.2$\pm$5.8 & 0.6$\pm$0.4 & 100 & 0 & 0 \\
 & 320 & 14$\pm$1.3 & 11$\pm$3.9 & 0.2$\pm$0.3 & 100 & 0 & 0 \\
 & 640 & 16$\pm$0.6 & 12$\pm$3.3 & 0.3$\pm$0.2 & 100 & 0 & 0 \\ \bottomrule
%\multicolumn{2}{c|}{\textit{\textbf{\begin{tabular}[c]{@{}c@{}}Average\end{tabular}}}} & \textit{+17\%} & \textit{+2\%} & \textit{+83\%} & \textit{+115\%} & \textit{+36\%} & \textit{\textbf{0\%}} \\ \bottomrule
\end{tabular}
}
\end{table}
\section{DISCUSSION} \label{sec:disc}
The evaluation illustrates how ASA combines a pro-active submission scheduling with Per-stage's strategy to simultaneously minimize resource usage and waiting times.
ASA can be specially useful when a workflow has multiple large consecutive stages, where the impacts of waiting in a queue can overtake the usefulness of non-monolithic applications and workflows, represented by the Big Job allocations. %\devarshi{somewhere it says BigJob and here it says Big Job. Need to be consistent with naming (either with or without spaces)}
%, this feature was not tested in here since it would need additional resource isolation control mechanisms, and these are outside the scope of this paper.

Table~\ref{tab:results} demonstrates that ASA achieves makespans close to those of Big Job allocations, while using as little resources as Per-Stage allocation strategy.
Big Job strategy results in shorter makespans, but always end up in larger resource usage (core-hours). 
\systemname's Per-Stage \cite{ehpc} strategy results in best resource usage and worse makespans.
ASA simultaneously tackles both, with close-to-optimal makespans when compared to Big Job's and specially to Per-Stage (Figures 6-8), with best core-hour usage (Table 1). 
In real systems, job submission planning is key, and as explained in Section~\ref{workflows}.
To avoid both bad estimations and violating workflow's ordering constraints, ASA uses Slurm dependency features to link the various stages. 
Thus ASA shows no losses, except in ASA Naive (Figure 5a) which does not use such features. 
Although the architecture supports collocation of different workflow tasks in other's workflow allocations and resources (Figure \ref{asa:architecture}), we preferred not managing resource allocations among different workflow stages. 
For instance, the architecture allows task co-placement from different workflows to share a same resource like CPU.
As mentioned, this is supported by Mesos in a fine-grained manner as Mesos supports resource capacity scheduling constraints to be specified, like for example CPU utilization: if one task uses only up to  10\% of a CPU resource, Mesos can co-schedule additional tasks in the same CPU up to a global threshold is reached (e.g. 100\%).
Although task co-placement optimizes overall resource usage if done correctly, it may have direct impacts on the time limits set by users, besides workflow performance impacts.
As additional actions would need to be studied to safely support such proposition, we decided to not do it in this paper.

\begin{figure}
	\centering
	\includegraphics[width=0.8\linewidth]{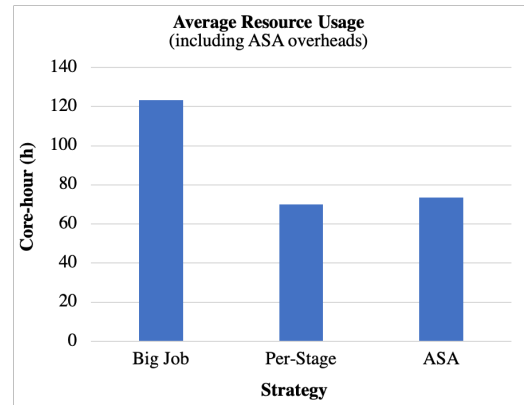}
	\caption{ASA - Total Resource usage for Montage, BLAST, and Statistics, in the three strategies: Big-Job, Per-Stage, and ASA. ASA overheads are included.}
	\label{asa:arch}
\end{figure}

Table \ref{tab:summary} shows how \systemone and \systemtwo affect ASA predictions and resource usage fares.
Although smaller jobs experience shorter queue waiting times, they experience variations of up to almost 1 hour in \systemone's queue workload, ASA controlled quite notably the core-hour overhead losses (OH).
A high queue variation negatively impacts ASA, causing its predictions to also vary largely during experiments because ASA has to adjust its probability distribution modelling the queue. 
For smaller job geometries ($\leq 112$ cores), ASA has to acquire knowledge from a large number $m$ (see previous Section) of alternatives until it can build-up knowledge for making accurate estimates.
However, as the system's load varies aggressively, ASA has to adapt to such variations to bound the overhead losses.
ASA is still able to reduce the perceived waiting times seen by workflows most of the time.
The high variation in a queue usually happens due to fragmentation caused in the system by smaller job geometries with varied similar, but not identical constraints, something that larger jobs ($\geq 160$ cores) in \systemtwo do not experience, and explains the high ASA accuracy in such system.

As explained in Section~\ref{sec:results}, ASA can be tuned to follow closely the last observed queue waiting time, which would change the results seen in Table \ref{tab:summary}, though its effects should be extensively studied in more specific scenarios.
Modern schedulers like Slurm allow dependencies to be set among different jobs, and such features would mitigate the core-hours overhead caused by over-estimations, as can be seen in Table~\ref{tab:results}.
Although job-dependencies enable overhead control, it may affect perceived queue waiting times because schedulers postpone job submissions until their dependencies are set.
Our experimental results from Table \ref{tab:summary} show that ASA can be specially useful for large job geometries (achieving 100\% accuracy), which can enable the resource planning capability as a feature.

Generally, our paper focuses on the first of two ASA features: to the best of our knowledge, a new, convergence proven (see Appendix A) Reinforcement-Learning method for estimating queue waiting times (WT) exclusively from user's perspective; a library for finer-grain management and scheduling of workflow's tasks (Mesos).
Rather than using traces and/or resource manager's queue waiting time estimates (which can speed-up ASA convergence), we opted for real experiments in production systems.
In this way we can evaluate how ASA would work as a general scheduling algorithm, and not only as a neat library enabling a diverse set of scheduling strategies to modern HPC systems.
Results summarize experiments using 1000s of core-hours across two production HPC systems with large differences in architecture, users, workload, etc (Table \ref{tab:summary}).
Algorithm \ref{alg.asa} is a very simple method which adapts its knowledge and estimations by adjusting mini-batches (or rounds), resetting them when bad estimates are detected so to bound its losses.
By sharing this information in a per job-geometry basis across different experimental scales, improvements in both systems are reported (Figures 5-8).
\section{CONCLUSIONS} \label{sec:conc}

%User experience with regards to High Performance Computing (HPC) system has been always attached to the perceived time to results.
%Although the idea of combining different scheduling techniques and novel frameworks on a single library has been proposed before, adaptive mechanisms have not been considered as a key aspect of such propositions.
%In fact, k
Ensuring optimal application performance for timely processing without resource wastage remains a persistent challenge, especially in systems tailored for low latency and highly dynamic data-intensive workflows. 
Addressing these challenges necessitates innovative solutions that combine classical approaches with adaptive, elastic, and fault-tolerant architectural features. 
To tackle these, and leveraging on user perceived system's performance, we propose ASA: the Adaptive Scheduling Algorithm. 
ASA learns and estimates the queue waiting times by using a novel reinforcement learning algorithm, which combined with its resource manager layer provides applications with the ability to dynamically adjust job resource planning based on workflow stage requirements. 
These allocations are done proactively based on the waiting time predictions to ensure that resources for subsequent workflow stages are available upon completion of ongoing stages.  
%The evaluation based on three real applications running with different configurations over two supercomputers demonstrates that ASA achieves makespans close to those of overallocated large jobs, with only 2\% higher than large allocations across all three workflows.
Our evaluation involving three real applications operating with diverse configurations across two supercomputers showcases that ASA attains makespans comparable to overallocated submissions, exhibiting a mere 2\% increase in total runtime.
Adopting a per-stage allocation approach, ASA achieves near-optimal total core-hour usage, representing a 43\% reduction compared to large allocations across all evaluated scenarios. Furthermore, ASA attains 100\% accuracy for large jobs, demonstrating resilience against high queue workload variability. Future work will focus on extending ASA with statefulness to support different metrics, accommodate heterogeneity, and enable more sophisticated proactive scheduling techniques, including multi-constraint/dimensional scheduling support.
%\input{appendix.tex}
%%
%% The acknowledgments section is defined using the "acks" environment
%% (and NOT an unnumbered section). This ensures the proper
%% identification of the section in the article metadata, and the
%% consistent spelling of the heading.
\begin{acks}
%To Robert, for the bagels and explaining CMYK and color spaces.
We are thankful for the support of the Swedish National Infrastructure for Computing (SNIC) and the Office of Advanced Scientific Computing Research (ASCR) of the U.S. Department of Energy under Contract Number DE-AC02-05CH11231.
\end{acks}

%%
%% The next two lines define the bibliography style to be used, and
%% the bibliography file.
\bibliographystyle{ACM-Reference-Format}
\bibliography{main.bib}

%\addtolength{\textheight}{-12cm}   % This command serves to balance the column lengths
                                  % on the last page of the document manually. It shortens
                                  % the textheight of the last page by a suitable amount.
                                  % This command does not take effect until the next page
                                  % so it should come on the page before the last. Make
                                  % sure that you do not shorten the textheight too much.

%%
%% If your work has an appendix, this is the place to put it.
%\section*{APPENDIX}
\appendix\label{asa:appendix}

\section{Convergence of ASA}

In this appendix we mathematically prove ASA's convergence towards the true waiting time, as shown in Figure~\ref{asa:convergence}.

\begin{Theorem}

	Let $\theta=(\theta_1, \dots, \theta_m)\in\mathbb{R}^m$ be a fixed, given collection of waiting time alternatives amongst which to choose.
 	Let the ASA algorithm run on a sequence of $t$ processes, 
	and let $\eta(t)$ denote the number of mini-batches created by the algorithm as of time $t$.
	Then for any $\delta>0$ with probability exceeding $1-\delta$, one has that 
	
	\begin{equation}
		\sum_{s=1}^t \ell_s\left(\theta^{s-1}\right) 
		- 
		\sum_{s=1}^t \ell_s(\bar\theta)
		\leq 
		4\eta(t)+ \ln(m) +  \sqrt{2t \ln\left(\frac{m}{\delta}\right)}.
		\label{eq.thm1.}
	\end{equation}
\end{Theorem}

\begin{proof}
	The key to this proof is to consider two different timescales:
	(1) runs from $1, \dots, t$ in a linear fashion, and 
	(2) runs over the same range in a different fashion as follows.
	Let $m_k\subset\{1, \dots,t\}$ such that each $m_k = a,\dots,b$  and $\cup_k m_k=\{1, \dots,t\}$.
	We refer to $m_k$ as a mini-batch, or \textit{round}, of length $|m_k|$.
	Consider a sequence $\{a_i, \dots, a_t\}$ for any $t$, then
	
	\begin{equation}
		\sum_{s=1}^t a_s 
		= 
		\sum_{k= 1}^{\eta(t)} \sum_{j\in m_k} a_{j},
		\label{eq.thm1.}
	\end{equation}
	
	with $\eta(t)$ the number of mini-batches $\{m_k\}$.
	
	Let $(s)$ point to the last {\em completed} mini-batch $m_k$ before iteration $s$.
	Rather than fixing the length of the mini-batches, the algorithm itself constructs the minibatches according to how well the learned solution is working.
	This extra layer of adaptivity enables the non-stationary setting.
	
	Let $\theta^{s-1}$ denote the estimated waiting time, randomly sampled according to $\textbf{p}_{(s-1)}$, 
	which is implemented for process $y_s$, with $s=1,2,3, \dots, t$.
	Define $Z_t>0$ as 
	\begin{multline}
		Z_t = \sum_{\theta_i} e^{- \sum_{s=1}^t  I\left(\theta_i=\theta^{s-1}\right) \ell_s(\theta_i)} 
		=
		\sum_{\theta_i} e^{-   \sum_{k=1}^{n^k} \sum_{j\in m_k}  I\left(\theta_i=\theta^{j-1}\right)\ell_j(\theta_i)}.
		\label{eq.thm1.}
	\end{multline}
	
	Then
	
	\begin{multline}
		\ln\frac{Z_t}{Z_0} = \ln(Z_t) - \ln(Z_0)
		= 
		\ln \sum_{\theta_i} e^{- \sum_{s=1}^t  I\left(\theta_i=\theta^{s-1}\right)  \ell_s(\theta_i)}
		-
		\ln(m) \\
		\geq 
		- \sum_{s=1}^t   I\left(\bar\theta=\theta^{s-1}\right) \ell_s(\bar\theta)
		-
		\ln(m).
		\label{eq.thm1.}
	\end{multline}
	
	Conversely,
	
	\begin{multline}
		\ln\frac{Z_{\bar{k}}}{Z_{\bar{k}-1} } 
		= 
		\ln\frac
		{\sum_{\theta_i} e^{- \sum_{k=1}^{\bar{k}} \sum_{j\in m_k}   I\left(\theta_i=\theta^{j-1}\right) \ell_j (\theta_i)}}
		{\sum_{\theta_i} e^{-\sum_{k=1}^{\bar{k}-1} \sum_{j\in m_k}   I\left(\theta_i=\theta^{j-1}\right)  \ell_j(\theta_i)}} 
		=
		\ln \sum_{\theta_i} \textbf{p}_{\bar{k}-1,i}  e^{\left( - \ell_{(\bar{k})}(\theta_i) \right)},
		\label{eq.thm1.}
	\end{multline}
	
	where we use the definition
	
	\begin{equation}
		\ell_{(\bar{k})}(\theta_i)
		\triangleq
		\sum_{j\in m_k}  I\left(\theta_i=\theta^{j-1}\right) \ell_j(\theta_i).
		\label{eq.thm1.}
	\end{equation}	
	
	Then using the inequality property $1-x\leq e^{-x}\leq 1-x+x^2$ for all $x\geq -1$, gives 
	
	\begin{multline}
		\ln \sum_{\theta_i} \textbf{p}_{\bar{k}-1,i}  e^{- \ell_{(\bar{k})} (\theta_i)}
		\leq
		\ln  \left(1 -  \sum_{\theta_i} \textbf{p}_{\bar{k}-1,i}  \ell_{(\bar{k})} (\theta_i)  +  \left(\sum_{\theta_i} \textbf{p}_{\bar{k}-1,i} \ell_{(\bar{k})}(\theta_i) \right)^2 \right) \\
		\leq
		\ln e^{-  \sum_{\theta_i} \textbf{p}_{\bar{k}-1,i}  \ell_{(\bar{k})}(\theta_i)  +  1} \\
		=
		-  \sum_{\theta_i} \textbf{p}_{\bar{k}-1,i}  \ell_{(\bar{k})}(\theta_i)  + 1,
		\label{eq.thm1.}
	\end{multline}
	
	where by construction $\left(\sum_{\theta_i} \textbf{p}_{\bar{k}-1,i} \ell_{(\bar{k})}(\theta_i) \right)^2\leq 4$ for any $\bar{k}$.
	
	In conclusion,
	
	\begin{multline}
		- \sum_{s=1}^t  I\left(\theta_i=\theta^{s-1}\right) \ell_s(\bar\theta)
		-
		\ln(m) \\
		\leq 
		\ln\frac{Z_t}{Z_0}
		= 
		\sum_{s=1}^t \ln\frac{Z_s}{Z_{s-t}} \\
		\leq 
		- \sum_{s=1}^t \sum_{\theta_i} \textbf{p}_{(s),i}   I\left(\theta_i=\theta^{s-1}\right)  \ell_s(\theta_i)
		+ 4\eta(t),
		\label{eq.thm1.}
	\end{multline}
	
	or
	
	\begin{equation}
		\sum_{s=1}^t \sum_{\theta_i} \textbf{p}_{(s),i}   I\left(\theta_i=\theta^{s-1}\right)  \ell_s(\theta_i)
		- 
		\sum_{s=1}^t \ell_s(\bar\theta)
		\leq 
		 4\eta(t) + \ln(m).
		\label{eq.thm1.}
	\end{equation}
	
	So by defining the expectation at iteration $s$ as 
	
	\begin{equation}
		\mathbb{E}_{s}[\cdot] = \sum_{\theta_i} \textbf{p}_{(s),i}  \ell_\cdot(\theta_i),
		\label{eq.thm1.}
	\end{equation}
	
	one gets
	
	\begin{equation}
		\sum_{s=1}^t\mathbb{E}_{s}[\ell_s(\theta^{s-1})]
		- 
		\sum_{s=1}^t \ell_s(\bar\theta)
		\leq 
		 4\eta(t) + \ln(m).
		\label{eq.thm1.}
	\end{equation}
	
	Finally, invoking Azuma's inequality \cite{chung2006concentration} gives 
	that with probability exceeding $1-\delta<1$, one has 
	
	\begin{equation}
		\sum_{s=1}^t \ell_s\left(\theta^{s-1}\right) 
		- 
		\sum_{s=1}^t \ell_s(\bar\theta)
		\leq 
		4\eta(t) + \ln(m) +  \sqrt{2t \ln\left(\frac{m}{\delta}\right)},
		\label{eq.thm1.}
	\end{equation}
	
	as desired.
\end{proof}

%Explain how different sampling policies affect the decision making of ASA.
%How does our strategy behaves when different "tunings" of the algorithm are used. How do they affect the estimation accuracy/quality.
%What happens when resources get ready for use before they are actually needed: impacts on makespan and total resource usage.
%Add some errors to your predictions in order to quantify these.
%Combination of different tunings and policies. Replay runs, collect data, make a table, and compare them.
%Think about a job re-submission policy.

%Figure \ref{asa:tests} shows how quickly ASA converges overtime to the true queue waiting time (dashed blue line), after multiple interactions.
%In order to test ASA's adaptability capabilities, the true waiting time was randomly varied at four occasions (iterations number 200, 400, 600, and 800).
%The default ASA (Algorithm \ref{alg.asa}), takes a while to perceive the changes and converge, but with a tuned ASA, where the perceived loss is reinforced into Algorithm \ref{alg.asa} (Policy Tuned), the algorithm starts to converge more quickly.
%A greedy approach is show in the dashed red line, where the minimum perceived loss is used to estimate the queue waiting times.
%Because we use a simple loss function (see Section IV), when the true waiting time suddenly drops, the greedy policy reaches a local minimum and does not behave correctly.

\end{document}